\documentclass[openany]{nature}
\usepackage[T1]{fontenc}
\usepackage[left]{lineno}
\usepackage{amsthm,amsmath,amssymb}
\usepackage{mathrsfs}
\usepackage{overpic}
\usepackage{multirow}
\usepackage[T1]{fontenc}
\usepackage{graphicx}
\usepackage{pdflscape}
\usepackage{hyperref}
\usepackage{threeparttable}
\usepackage{threeparttablex}
\usepackage{xcolor}
\usepackage{ulem}
\usepackage{cancel}
\usepackage[figuresleft]{rotating}
\usepackage{caption}
\usepackage{subfigure}
\usepackage{tablefootnote}
\usepackage{hyperref}
\usepackage{graphicx}
\usepackage{longtable}
\usepackage{supertabular}
\usepackage{float} 
\usepackage{authblk}
\usepackage{multibib}
\usepackage{subfiles}
\usepackage{enumerate}
\usepackage{bm}
\usepackage{booktabs}
\usepackage{rotating}
\usepackage{longtable}
\usepackage{subcaption}
\usepackage[dvipsnames]{xcolor}
\usepackage{multicol}
\usepackage{enumitem}
\usepackage{subfiles}
\newcites{methods}{References}

\makeatletter
\renewcommand{\@cite}[2]{[{#1\if@tempswa , #2\fi}]}
\makeatother

\hypersetup{
colorlinks=true,
linkcolor=red,
filecolor=blue,
urlcolor=cyan,
citecolor=green}

\makeatletter
 
\def\emph#1 {\textit{ #1 } }

\let\saved@includegraphics\includegraphics
\AtBeginDocument{\let\includegraphics\saved@includegraphics}
\renewenvironment*{figure}{\@float{figure}}{\end@float}
\makeatother

\definecolor{dkblue}{RGB}{54, 86, 169}

\newcommand{\comment}[1]{}

\def\be{\begin{eqnarray}}
\def\ee{\end{eqnarray}}

\makeatletter

\title{Minutes-long soft X-ray prompt emission from a compact object merger}

\author{An Li$^{1,2}$\thanks{These authors contributed equally to this work\\Corresponding authors:}, 
Chen-Wei Wang$^{3,4*}$,
Niccol\`o Passaleva$^{5,6*}$, 
Jie An$^{4,7*}$,
Bin-Bin Zhang$^{8,9}$\thanks{E-mail: bbzhang@nju.edu.cn}, 
Eleonora Troja$^{6,10}$\thanks{E-mail: eleonora.troja@uniroma2.it}, 
Yi-Han Iris Yin$^{11,12}$\thanks{E-mail: iris.yh.yin@connect.hku.hk},
Yuan Liu$^{7}$, 
Shao-Lin Xiong$^{3}$, 
Li-Ping Xin$^{7}$, 
Yi-Xuan Shao$^{8,9}$, 
Jun Yang$^{13}$, 
Hui Sun$^{7}$,
Dong Xu$^{7}$, 
Yu-Han Yang$^{6}$, 
Roberto Ricci$^{6,14}$, 
He Gao$^{1,2}$, 
Sarah Antier$^{15}$,
Rosa L. Becerra$^{6,16}$, 
Jia-Xin Cao$^{17}$, 
Alberto Javier Castro-Tirado$^{18}$, 
Xin-Lei Chen$^{19}$,
Ye-Hao Cheng$^{19}$, 
Yong Chen$^{3}$, 
Hua-Qing Cheng$^{7}$, 
Valerio D'Elia$^{20}$,
Massimiliano De Pasquale$^{21}$, 
Yong-Wei Dong$^{3}$, 
Eslam Elhosseiny$^{22}$,
Rob A. J. Eyles-Ferris$^{23}$, 
Maria Gritsevich$^{24,25}$, 
Xu-Hui Han$^{7}$, 
Dieter Hartmann$^{26}$,
You-Dong Hu$^{17}$, 
Jing-Wei Hu$^{7}$\thanks{E-mail: hujingwei@nao.cas.cn}, 
Shu-Mei Jia$^{3}$, 
Nino Kochiashvili$^{27}$,
Wei-Hua Lei$^{28}$, 
Andrew J. Levan$^{29,30}$, 
Cheng-Kui Li$^{3}$, 
Dong-Yue Li$^{7}$, 
Hua-Li Li$^{7}$\thanks{E-mail: lhl@nao.cas.cn},
Xiao-Bo Li$^{3}$, 
Zhi-Xing Ling$^{4,7}$, 
He-Yang Liu$^{7}$, 
Hou-Jun L\"{u}$^{17}$, 
Daniele B. Malesani$^{31,32,29}$,
Brendan O'Connor$^{33}$, 
Hai-Wu Pan$^{7}$, 
Shashi Bhushan Pandey$^{34}$, 
Ignacio Perez-Garcia$^{35}$,
Dani\"elle L.~A. Pieterse$^{29}$, 
Marion Pillas$^{36}$, 
Yu-Lei Qiu$^{7}$, 
Andrea Saccardi$^{37}$,
Rub\'en S\'anchez-Ram\'irez$^{38}$, 
Wen-Jun Tan$^{3,4}$, 
Manasanun Tanasan$^{39}$, 
Nial R. Tanvir$^{23}$,
Susanna D. Vergani$^{40}$, 
Jing Wang$^{7}$, 
Xiao-Feng Wang$^{41}$, 
Qin-Yu Wu$^{4,7}$, 
Shu-Xu Yi$^{3}$,
Tillayev Yusufjon$^{42,43}$, 
Chen Zhang$^{7}$, 
Wen-Da Zhang$^{7}$, 
Yi-Jia Zhang$^{41}$, 
Guo-Ying Zhao$^{44}$,
Chao Zheng$^{3}$, 
Shi-Jie Zheng$^{3}$, 
Chang Zhou$^{28}$, 
Ping Zhou$^{8}$, 
Bertrand Cordier$^{45}$,
Jian-Yan Wei$^{7}$, 
Weimin Yuan$^{4,7}$, 
Shuang-Nan Zhang$^{3,4}$, 
Bing Zhang$^{11,12,46}$
}

\begin{document}

\maketitle
\begin{affiliations}

\item School of Physics and Astronomy, Beijing Normal University, Beijing 100875, China
\item Institute for Frontier in Astronomy and Astrophysics, Beijing Normal University, Beijing 102206, China
\item State Key Laboratory of Particle Astrophysics, Institute of High Energy Physics, Chinese Academy of Sciences, Beijing 100049, China
\item University of Chinese Academy of Sciences, Chinese Academy of Sciences, Beijing 100049, China
\item Department of Physics, University of Rome "Sapienza", Rome, Italy
\item Department of Physics, University of Rome "Tor Vergata", Rome, Italy
\item National Astronomical Observatories, Chinese Academy of Sciences, Beijing 100101, China
\item School of Astronomy and Space Science, Nanjing University, Nanjing 210023, China
\item Key Laboratory of Modern Astronomy and Astrophysics, Nanjing University, Ministry of Education, Nanjing 210023, China
\item INAF, Istituto di Astrofisica e Planetologia Spaziali, Rome, Italy
\item The Hong Kong Institute for Astronomy and Astrophysics, The University of Hong Kong, Hong Kong, China
\item Department of Physics, The University of Hong Kong, Hong Kong, China
\item Institute for Astrophysics, School of Physics, Zhengzhou University, Zhengzhou 450001, China
\item INAF, Istituto di Radioastronomia, Bologna, Italy
\item IJCLab, Univ Paris-Saclay, CNRS/IN2P3, Orsay, France
\item Instituto de Astronom{\'\i}a, Universidad Nacional Aut\'onoma de M\'exico, Ciudad de M\'exico, M\'exico
\item Guangxi Key Laboratory for Relativistic Astrophysics, Department of Physics, Guangxi University, Nanning 530004, China
\item Instituto de Astrof\'isica de Andaluc\'ia (IAA-CSIC), Glorieta de la Astronom\'ia s/n, E-18008, Granada, Spain
\item School of Physics and Astronomy, Yunnan University, Kunming 650500, China
\item Space Science Data Center (SSDC) - Agenzia Spaziale Italiana (ASI), Via del Politecnico snc, Roma 00133, Italy
\item Department of Mathematics and Computer Sciences, Physical Sciences and Earth Sciences of University of Messina, Via F.S. D'Alcontres 31, Messina 98166, Italy
\item National Research Institute of Astronomy and Geophysics (NRIAG), 1 El-marsad St., 11421 Helwan, Cairo, Egypt
\item School of Physics and Astronomy, University of Leicester, University Road, Leicester LE1 7RH, UK
\item Faculty of Science, University of Helsinki, Gustaf Hallströmin katu 2, Helsinki FI-00014, Finland
\item Institute of Physics and Technology, Ural Federal University, Mira str. 19, Ekaterinburg 620002, Russia
\item Clemson University, Department of Physics and Astronomy, Clemson SC 29634-0978, USA
\item E. Kharadze Georgian National Astrophysical Observatory, Mt. Kanobili, Abastumani, Adigeni 0301, Georgia
\item Department of Astronomy, School of Physics, Huazhong University of Science and Technology, Wuhan 430074, China
\item Department of Astrophysics/IMAPP, Radboud University Nijmegen, P.O.~Box 9010, Nijmegen 6500~GL, The Netherlands
\item Department of Physics, University of Warwick, Coventry CV4~7AL, UK
\item Cosmic Dawn Center (DAWN), Denmark
\item Niels Bohr Institute, University of Copenhagen, Jagtvej~155, 2200~Copenhagen~N, Denmark
\item McWilliams Center for Cosmology and Astrophysics, Department of Physics, Carnegie Mellon University, Pittsburgh, PA, USA 
\item Aryabhatta Research Institute of Observational Sciences (ARIES), Manora Peak, Nainital, Uttarakhand 263001, India
\item Instituto de Astrof\'isica de Andaluc\'ia (IAA-CSIC), Granada, Spain
\item Sorbonne Université, CNRS, UMR 7095, Institut d'Astrophysique de Paris, 98 bis boulevard Arago, Paris 75014, France
\item Universit\'e Paris-Saclay, Universit\'e Paris Cit\'e, CEA, CNRS, AIM, 91191, Gif-sur-Yvette, France
\item Instituto de Astrof\'{\i}sica de Andaluc\'{\i}a, Consejo Superior de Investigaciones Cient\'{\i}ficas (IAA-CSIC), Glorieta de la Astronom\'{\i}a, s/n, Granada 18080, Spain
\item National Astronomical Research Institute of Thailand (NARIT), Chiang Mai 50180, Thailand
\item LUX, Observatoire de Paris, Universit\'e PSL, CNRS, Sorbonne Universit\'e, Meudon 92190, France
\item Department of Physics, Tsinghua University, Beijing 100084, China
\item Ulugh Beg Astronomical Institute, Uzbekistan Academy of Sciences, Astronomy St. 33, Tashkent 100052, Uzbekistan
\item National University of Uzbekistan, University Str. 4, Tashkent 100174, Uzbekistan
\item School of Physics and Astronomy, Sun Yat-Sen University, Zhuhai, 519082, China 
\item CEA Paris-Saclay, Irfu/Département d'Astrophysique, 9111 Gif sur Yvette, France
\item Nevada Center for Astrophysics and Department of Physics and Astronomy, University of Nevada Las, Vegas, NV 89154, USA
\end{affiliations}

\begin{abstract}
{\bf Abstract\\}
\normalfont
Compact object mergers are multi-messenger sources and known progenitors of some gamma-ray bursts \cite{2017ApJ...848L..13A,1989Natur.340..126E}, bright flashes of high-energy radiation powered by a central engine, either an accreting black hole or a neutron star. Our understanding of these events has so far been shaped primarily by observations in the gamma-ray band \cite{1993ApJ...413L.101K,2006ApJ...643..266N}, leaving their prompt phase poorly constrained at lower energies \cite{2005Natur.437..855V}. A long-lasting ($\approx$100 s) engine-driven X-ray emission was discussed \cite{2013ApJ...763L..22Z, 2019ApJ...886..129S} to explain rapidly fading X-ray afterglows \cite{2013MNRAS.430.1061R,2015ApJ...805...89L} following several ($\approx$30\%) bursts of short ($\lesssim$2 s) duration. However, this prompt X-ray component was not directly observed  and past candidates were not confirmed \cite{2019Natur.568..198X, 2025A&A...695A.279Q}. Here we report the discovery of EP250704a containing a minutes-long ($\sim$560 s) flash of soft (0.5--4 keV) X-rays immediately following the short ($\sim$0.4 s) GRB 250704B. The variability and spectral shape of this emission are inconsistent with the canonical picture of a hard, accretion-powered spike followed by a standard external-shock afterglow \cite{2005Natur.437..855V,1997ApJ...476..232M}. Instead, the long-soft bump points to a distinct phase of prompt emission in X-rays, which would not have been detected without the soft X-ray coverage of Einstein Probe. The detection of a prompt soft X-ray counterpart in an otherwise ordinary short GRB shows that long-lasting X-ray emission is likely a common feature of merger-driven bursts and a promising electromagnetic counterpart to gravitational wave sources.
\end{abstract}

\bigskip
\noindent{\bf keywords:} Fast X-ray transient; Compact object merger; Gamma-ray burst; Magnetar; Tianguan Einstein Probe

\section{Introduction}

Canonical GRBs have been studied primarily through their gamma-ray emission \cite{1993ApJ...413L.101K,2006ApJ...643..266N, 2018pgrb.book.....Z}. 
Recent observations by the Einstein Probe (EP) mission have shown that they also exhibit a rich phenomenology at lower energies, including precursors and long-lasting tails of soft X-ray emission  \cite{2024ApJ...975L..27Y, 2025NatAs...9..564L, 2025ApJ...988L..34J, Yin_2025}. 
However, to date, these observations were limited to the class of long duration GRBs, for which a long-lived central engine is a natural requirement capable of accommodating their observed properties at gamma-ray and X-ray energies. 

Prompt X-ray observations of short duration GRBs remained instead exceedingly rare \cite{2005Natur.437..855V}. Short GRBs, lasting less than two seconds in the gamma-ray band, are now firmly linked to compact object mergers \cite{2017ApJ...848L..13A, 1989Natur.340..126E,2017ApJ...848L..12A}. 
Although a temporally extended $\gamma$-ray  emission was detected in a small subset of events \cite{2006ApJ...643..266N, 2013ApJ...763L..22Z, 2019ApJ...886..129S}, 
the short duration measured in the majority of the sample continued to support a short-lived central engine, as expected in standard accretion-powered models. 
In this paper, we report the discovery of a fast X-ray transient, EP250704a, detected by the EP mission \cite{2022hxga.book...86Y}, together with its temporal and spatial association with the short GRB 250704B observed by Space Variable Objects Monitor (SVOM)/Gamma-ray Burst Monitor (GRM) \cite{2025GCN.40940....1S} and Insight-HXMT \cite{2025GCN.40978....1W}. The multi-episode X-ray emission properties, including rapid variability and spectral evolution, point to a distinct phase of long, soft prompt emission accompanying an otherwise short duration GRB. 
Multiwavelength afterglow behaviors and stringent constraints from the non-detection of an associated supernova, collectively suggest prolonged central engine activity following a compact binary merger. 

\section{Observations and Data Analysis}

EP250704a is a long-lasting X-ray transient discovered in the 0.5--4\,keV band by the Wide-field X-ray Telescope (WXT) onboard the EP mission \cite{2022hxga.book...86Y}. It is detected starting at $\sim T_0+0.2\,\mathrm{s}$ \cite{2025GCN.40941....1L}, where $T_0=08{:}16{:}27.10$ UTC on 2025 July 4 denotes the trigger time of GRB~250704B, a short duration ($T_{90}=0.37\pm0.06$ s; Table \ref{tab:obs_properties}) GRB independently discovered by SVOM/GRM \cite{2025GCN.40940....1S}, Insight-HXMT \cite{2025GCN.40978....1W}, and Konus-Wind \cite{2025GCN.40972....1F} (Methods). 
The WXT light curve of EP250704a, shown in Fig.~\ref{fig:detection} and Fig.~\ref{fig:EP250704a}a,  comprises three episodes separated by periods of apparent quiescence: an initial sharp spike (Episode I) beginning at {$T_0+0.25$ s} with a total duration of $T_{90,\mathrm{X}}=0.35^{+0.01}_{-0.05}$ s (Fig.~\ref{fig:EP250704a}b); a seconds-long tail (Episode II), followed by a temporally extended X-ray bump (Episode III) of {$\sim540$ s} starting at {$\sim T_0+20$ s}. The initial spike of EP250704a is temporally and spatially coincident with the brightest pulse of GRB~250704B, unambiguously tying these two high-energy transients to a common origin \cite{2025GCN.41050....1K}.
However, their temporal profiles are drastically different: whereas EP250704a displays a long-lasting multi-episode prompt X-ray emission (Episode II and Episode III), no gamma-rays are detected during this phase down to $2.13\times10^{-9}~{\rm erg~cm^{-2}~s^{-1}}$ in the 15--150 keV energy range (Fig.~\ref{fig:detection}e).

Emission from EP250704a triggered WXT at $\sim T_0+25$ s, initiating an automatic slew with the Follow-up X-ray Telescope (FXT) \cite{2025GCN.40956....1L}. 
Starting {at $\sim T_0+227$ s}, FXT observations localized a fading X-ray afterglow  (Fig.~\ref{fig:detection}a, b). 
Subsequent optical and radio follow-up identified counterparts and established a redshift of $z=0.66102 \pm 0.00011$ (see Methods). 
At this distance, the prompt gamma-ray spectrum of  GRB~250704B/EP250704a peaks at $E_{\rm peak, z}=978_{-203}^{+153}$ keV (rest-frame) and the isotropic-equivalent energy released in the 10--1000 keV band is $E_{\rm iso}=3.79_{-0.25}^{+0.20}\times10^{51}$ erg. 
These values place the burst well outside the population of GRBs from collapsing massive stars \cite{2006MNRAS.372..233A} (collapsars; Type II) and within the population of merger-driven GRBs (Type I; Fig.~\ref{fig:correlations}). 
At $z=0.661$, the distinctive signals of a merger-driven explosion, either a gravitational wave chirp or an optical/infrared kilonova, are beyond the reach of current ground-based facilities. Nonetheless, our deep optical limits rule out a broad range of supernovae \cite{1998Natur.395..670G,1999Natur.401..453B}, making a collapsar origin unlikely (Methods). Therefore, beyond the short GRB duration itself, standard optical and gamma-ray constraints strengthen the association between EP250704a/GRB~250704B and a compact-binary merger.

Additionally, the properties of EP250704a itself distinguish it from the larger population of extragalactic fast X-ray transients (EFXTs). Although multiple-episode behavior is seen in some EFXTs, the presence of a short, well-defined leading spike (Episode I) in EP250704a is unprecedented among EFXTs detected by EP (Fig. \ref{fig:EP250704a}e). A Bayesian-block analysis of this feature yields a minimum variability timescale (MVT) of $\sim38$\,ms, an outlier relative to the EP sample (Fig. \ref{fig:EP250704a}c). On the MVT–$T_{90}$ diagram, both the initial X-ray spike of EP250704a and the gamma-ray emission of GRB~250704B fall within the region of Type I bursts (Fig. \ref{fig:EP250704a}d). Consistently with this result, we find that the sharp X-ray spike (Episode I) in EP250704a is morphologically and spectrally tied to the bright, second gamma-ray pulse in the short GRB~250704B ({Fig. \ref{fig:consistency}; Methods), and can be interpreted as its low-energy extension. 
In this context, the delayed X-ray onset arises naturally from the compactness and baryon loading of the emitting region during the initial  gamma-ray pulse: self-absorption suppresses the low-energy spectrum early on \cite{Shen2009}, but as the jet excavates a low-density funnel through the merger ejecta, later prompt episodes reach larger radii and emerge in X-rays \cite{2002ApJ...581.1236Z} (Methods).

A connection between compact binary mergers and long duration EFXTs has already been proposed \cite{2013ApJ...763L..22Z, 2019Natur.568..198X,2019ApJ...886..129S}, but never definitively confirmed in the canonical short GRB merger systems \cite{2025A&A...695A.279Q}. In an earlier example, a long-soft X-ray component was detected in the short GRB 050709. However it was interpreted as standard afterglow onset owing to its smoothness and lack of spectral evolution \cite{1997ApJ...476..232M, 2005Natur.437..855V}. 
In EP250704a, by contrast, multi-epoch, broadband coverage reveals clear signatures of ongoing central-engine activity, providing the missing link between some EFXTs and their merger progenitors.
Three observational facts point to an internal, engine-powered origin for the temporally extended X-ray component up to $\sim T_0+400$ s:
{(i) rapid variability with \(\Delta t/t\!\lesssim\!0.1\), incompatible with a forward-shock afterglow \cite{2005Sci...309.1833B,2006ApJ...642..354Z} (Fig.~\ref{fig:variability_plane}; Methods); (ii) a subsequent spectral hardening (Fig.~\ref{fig:fxt_spec_evo}; Methods)—robust whether absorption is fixed or free—marking the transition from prompt to afterglow emission \cite{Yin_2025}; and (iii) temporal and spectral indices comparable to early X-ray light curves of short GRBs (Fig.~\ref{fig:comparison}) attributed to internal plateaus \cite{2015ApJ...805...89L}.
Additionally, a long-lived central engine is supported by the afterglow evolution at later times. 
The shallow X-ray decay from $T_0+690$ s to $7.2\times 10^4$ s together with the slow optical rise from $T_0+2\times 10^3$ s to $6.5\times10^4$ s require sustained energy injection, which might be provided by a long-lived central engine (Fig.~\ref{fig:afterglow}; Methods). Therefore, we attribute the extended X-ray bump (Episode III) to long-lasting, engine-powered emission, indicating that the merger remnant—most likely a magnetar—remained active well beyond the gamma-ray phase. The X-ray emission may originate from internal dissipation of a relativistic, highly magnetized magnetar wind, as envisioned in \cite{2013ApJ...763L..22Z}, given that the detection of the short GRB implies a line of sight close to the jet axis.

\section{Discussion and Conclusion}

Historically, long-lasting prompt emission in merger-driven bursts has been identified only in a minority ($\simeq$ 10\%–15\%) of cases — short GRBs with extended emission \cite{2006ApJ...643..266N,2021ApJ...911L..28D,2008MNRAS.385.1455M} and rare long GRBs with hybrid properties \cite{2006Natur.444.1044G,2022Natur.612..228T, 2022Natur.612..232Y, 2025NSRev..12E.401S}.
By revealing an X-ray bump in GRB 250704B, EP observations shift this paradigm. 
EP250704a demonstrates that a merger-driven burst can look like an ordinary short duration GRB in gamma-rays while exhibiting minutes-long emission in soft X-rays. The presence of X-ray plateaus was indirectly inferred by catching their final decaying tails \cite{2013MNRAS.430.1061R,2015ApJ...805...89L} with the Neil Gehrels Swift Observatory, however, their onset, temporal profile and short-term variability were not observed due to the slew time gap.

Extrapolating the average flux of the soft X-ray bump to the 15--150 keV band places it below the detection threshold of the Burst Alert Telescope (BAT \cite{2005SSRv..120..143B}) aboard the Swift. Therefore, had EP250704a been monitored solely in gamma-rays, only the 0.4-s spike would have been detected. This is not merely a tip-of-the iceberg effect caused by the burst's distance scale. Detectability with Swift/BAT becomes feasible at $z\lesssim 0.08$, and even then the X-ray bump would be at best marginally visible near the $\sim10^{-8}\ {\rm erg~cm^{-2}~s^{-1}}$ threshold \cite{2005SSRv..120..143B}. 
This is not the typical distance scale of
short GRBs, localized by Swift at $z\gtrsim 0.1$ \cite{2022MNRAS.515.4890O}. 
A long-lasting prompt X-ray component might therefore be prevalent in the sample of cosmological short GRBs, remaining undetected in most cases (Methods).

\begin{table*}
\centering
\renewcommand\arraystretch{0.8}
\begin{threeparttable}
\caption{\textbf{Observational properties.} All errors represent the 1$\sigma$ uncertainties.}
\label{tab:obs_properties}
\begin{tabular}{lc}
\hline
Observed Properties & EP250704a/GRB~250704B \\
\hline
\textbf{Soft X-ray [0.5--4 keV]:} & \\
Total duration ($\rm s$) & $561.91_{-94.01}^{+76.38}$ \\
Spike duration ($\rm s$) & $0.35_{-0.05}^{+0.01}$ \\
EE spectral index $\alpha$ & $-2.30 \pm 0.23$ \\
EE flux ($\rm erg~cm^{-2}~s^{-1}$) & $2.16_{-0.19}^{+0.22} \times 10^{-10}$\\
EE fluence ($\rm erg~cm^{-2}$) & $1.61_{-0.14}^{+0.16} \times 10^{-7}$ \\
EE luminosity ($\rm erg~s^{-1}$) & $2.61_{-0.23}^{+0.27} \times 10^{47}$ \\ 
EE Isotropic energy ($\rm erg$) & $1.93_{-0.17}^{+0.20} \times 10^{50}$\\
\hline
\textbf{Gamma-ray [10--1000 keV]:} & \\
Trigger time & 2025-07-04 08:16:27.100 \\
Duration ($\rm s$) & $0.37\pm0.06$ \\
Effective amplitude & 1.48 $\pm$ 0.03 \\
Minimum variability timescale ($\rm ms$) & $5$ \\
Rest-frame spectral lag\tnote{*} ($\rm ms$) & 2.5 $^{+3.5}_{-2.5}$ \\
Spectral index $\alpha_1$ & ${-0.78}_{-0.27}^{+0.31}$ \\
Spectral index $\alpha_2$ & ${-1.46}_{-0.21}^{+0.22}$ \\
Break energy $E_{\rm b}$ ($\rm keV$) & ${55.85}_{-26.47}^{+57.13}$ \\
Peak energy $E_{\rm p}$ ($\rm keV$) & ${588.84}_{-122.18}^{+91.93}$ \\
Peak flux ($\rm erg~cm^{-2}~s^{-1}$) & $1.23_{-0.03}^{+0.03} \times 10^{-5}$ \\
Total fluence ($\rm erg~cm^{-2}$) & $3.15_{-0.21}^{+0.17} \times 10^{-6}$ \\
Peak luminosity ($\rm erg~s^{-1}$) & $1.80_{-0.07}^{+0.06} \times 10^{52}$ \\
Isotropic energy ($\rm erg$) & $3.79_{-0.25}^{+0.20} \times 10^{51}$ \\
\hline
\textbf{Afterglow:} & \\
Redshift & $0.66102\pm0.00011$ \\
\hline
\textbf{Associations:} & \\
Kilonova & Not detectable at z=0.661 \\
Supernova & Ruled out: SN2025kg, SN1998bw, SN2006aj \\
\hline
\end{tabular}
\begin{tablenotes}
\item [*] The rest-frame spectral lag is measured between rest-frame energy bands 100–150 and 200–250 keV.
\end{tablenotes}
\end{threeparttable}
\end{table*}

\begin{figure}
\centering
\includegraphics[width=1.0\linewidth]{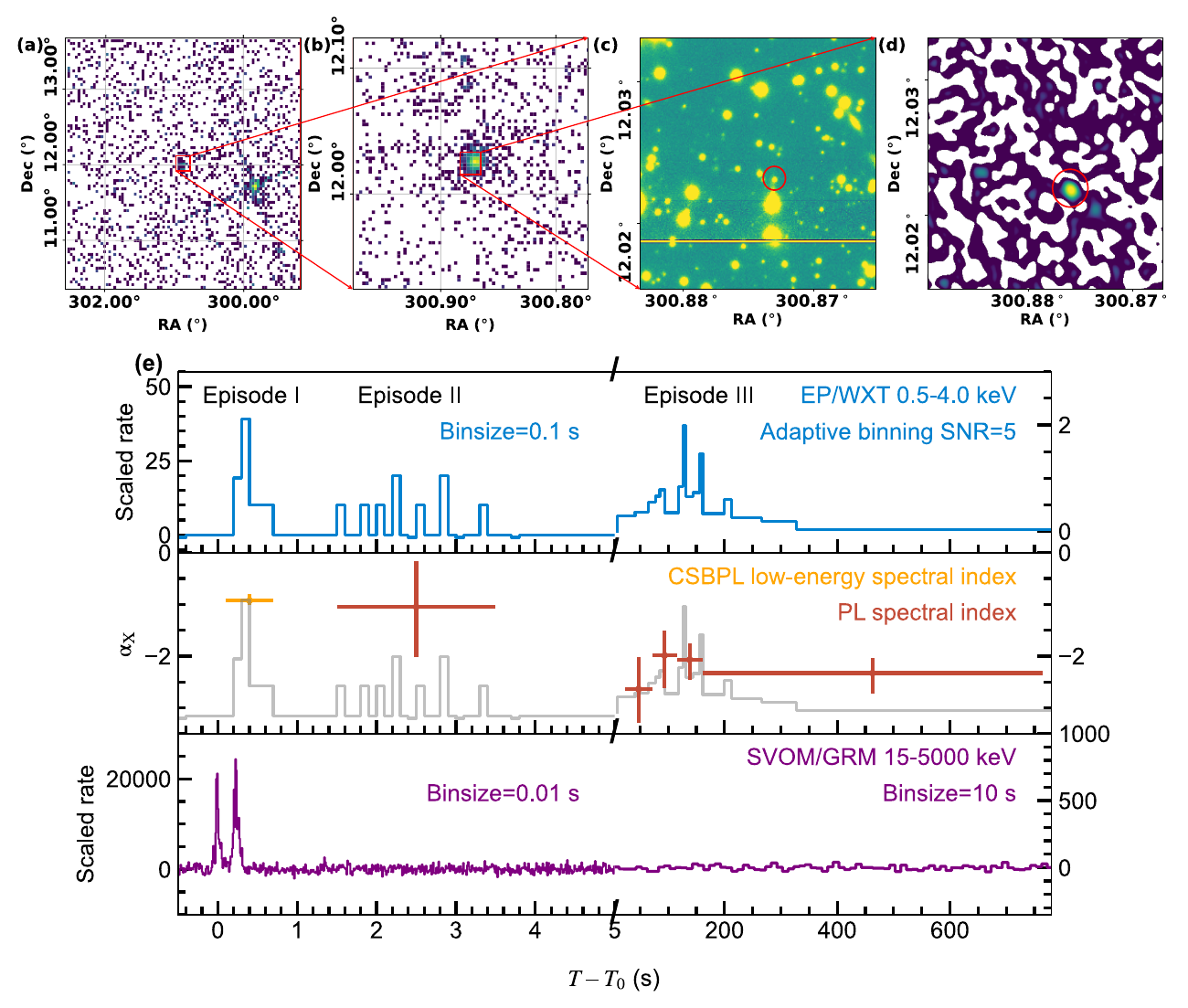}
\caption{
\textbf{Detection images and light curves for EP250704a/GRB~250704B in different bands.} 
(a) The image generated from the clean event data of EP-WXT, the GRB source is recognized with the red box.
(b) The image generated from the clean event data of EP-FXT, the GRB source is recognized with the red box.
(c), The image from VLT captured by FORS2, the optical afterglow of the burst is highlighted in the red circle.
(d) The image from VLA captured in 10 GHz, the radio afterglow of the burst is highlighted in the red circle.
(e) The light curves and spectral evolution of EP250704a/GRB~250704B. The blue and grey curves show the light curves of EP250704a in the energy range of 0.5--4.0 keV. The purple curve represents the light curve of GRB~250704B in the energy range of 15--5000 keV. The yellow and red points denote the low-energy spectral indices derived from the best-fit parameters of PL and CSBPL models, respectively. All error bars on data points indicate their 1$\sigma$ confidence level.
}
\label{fig:detection}
\end{figure}

\begin{figure}
\centering
\includegraphics[width=0.75\linewidth]{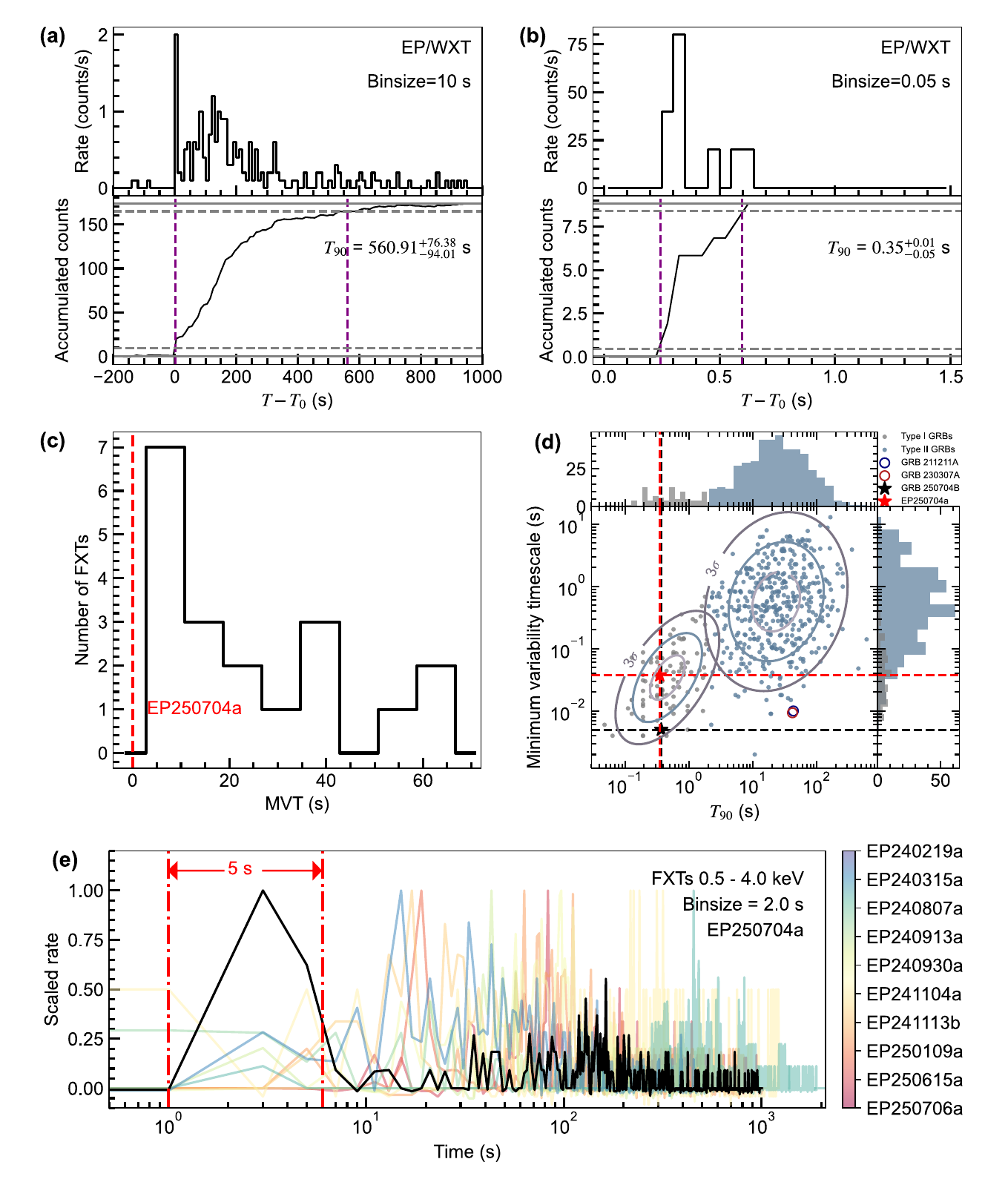}
\caption{
\noindent\textbf{Temporal and statistical analyses of EP250704a.}
(a) $T_{90}$ calculation of the EP/WXT detection of EP250704a. 
(b) $T_{90}$ calculation of the initial spike of EP250704a detected by EP/WXT. 
In (a) and (b), the black curves represent the light curve of EP/WXT in the energy range of 0.5--4 keV and the accumulated counts. The purple dashed vertical lines denote the $T_{90}$ interval. The error bars mark the 1$\sigma$ confidence level.
(c) Minimum variability timescale distribution of EP-detected FXTs. The red dashed vertical line denotes the location of EP250704a on the plot.
(d) The minimum variability timescale versus $T_{90}$ diagram. Type I and type II GRBs are represented by grey and cadet blue solid circles, respectively. EP250704a (Episode I) and GRB~250704B are highlighted by red and black stars, respectively.
(e) The scaled light curves of the well-sampled EFXTs detected by EP during 1.5 years of operation, shown with the bin size of 2 s. 
}
\label{fig:EP250704a}
\end{figure}

\begin{figure} 
\centering
\includegraphics[width=0.9\linewidth]{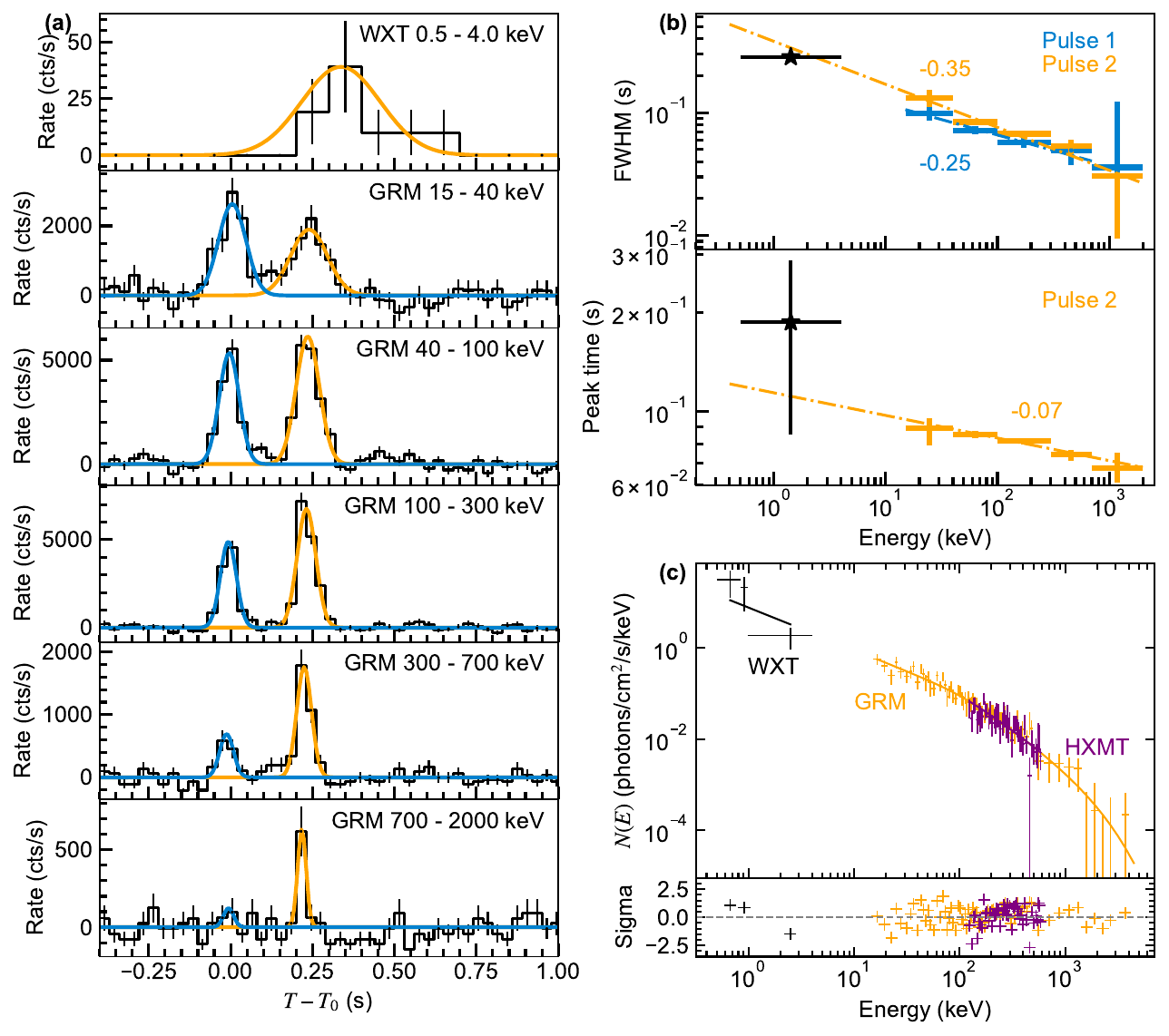}
\caption{
\noindent\textbf{Temporal and spectral consistency of the second pulse in the initial short spike.}
(a) Multiwavelength light curves of EP250704a/GRB~250704B. The WXT light curve is binned at 0.1 s, and the GRM light curves at 0.03 s. The blue and yellow curves are the best-fit Gaussian models for GRM light curves and an empirical Gaussian profile for the WXT light curve, respectively. 
(b) The FWHMs derived from the Gaussian models and pulse peak times as a function of energy. The peak times are relative to the starting time of the second gamma-ray pulse at $\sim 0.15$ s.
(c) The observed and modeled photon spectrum of the second pulse. All error bars mark the 1$\sigma$ confidence level.
}
\label{fig:consistency}
\end{figure}

\begin{figure} 
\centering
\includegraphics[width=1.00\linewidth]{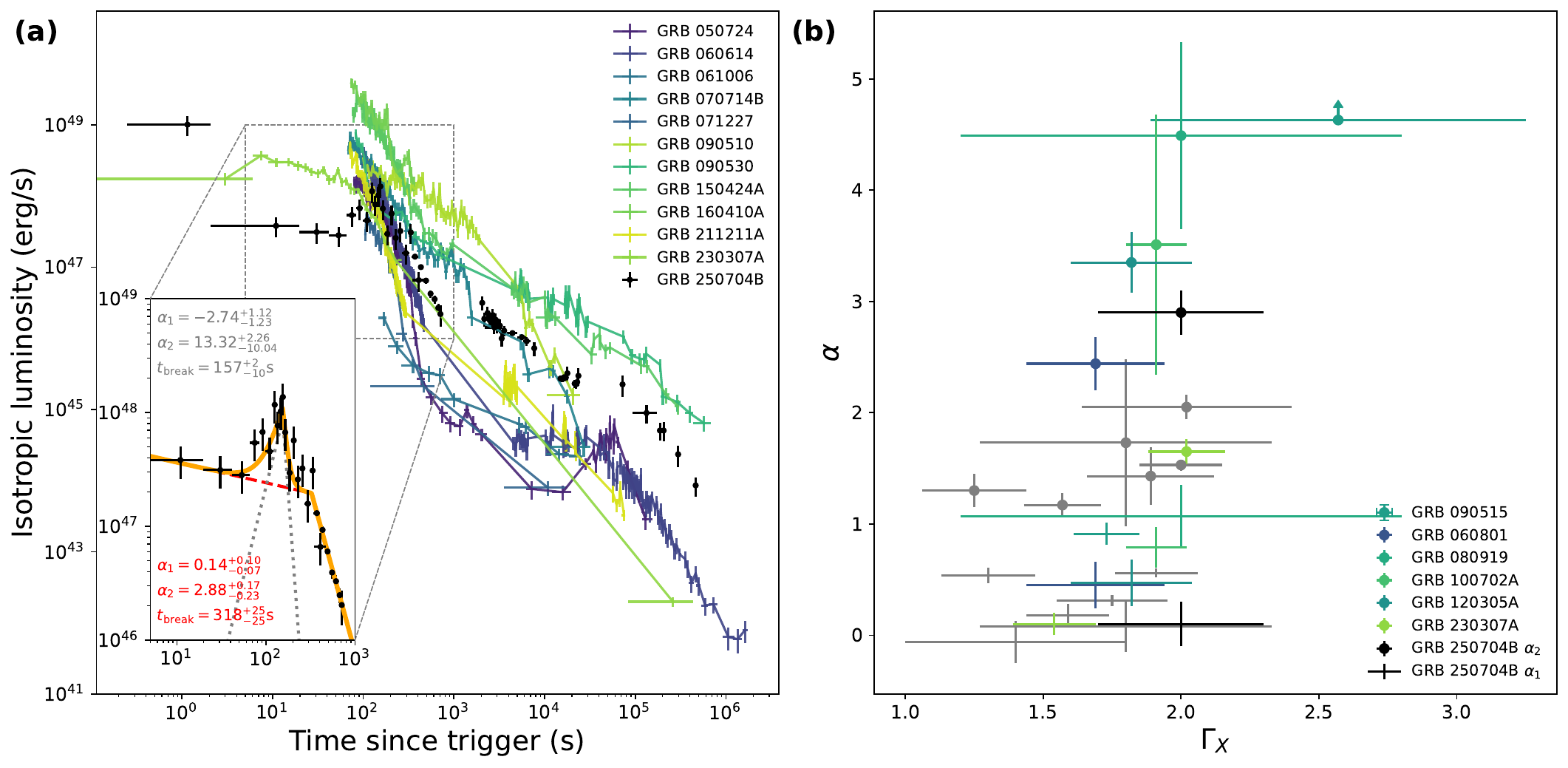}
\caption{\textbf{Comparison between GRB~250704B and other merger-driven GRB afterglows.} 
(a) A comparison of the X-ray luminosity light curves between GRB 250704B and a sample of merger-type GRBs with extended emission. The inset shows the simultaneous modeling of the plateau, steep decay and flare component using two broken power-law models.
(b) Photon index $\Gamma_X$ and decay-index comparison for the plateau phase ($\alpha_1$, small symbols) and the subsequent decay phase ($\alpha_2$, large symbols) for the sample of merger-driven GRBs with shallow decay. Grey dots show the normal shallow decay and colored dots show bursts with internal plateau.}
\label{fig:comparison}
\end{figure}

\begin{figure}
\centering
\includegraphics[width=1.0\linewidth]{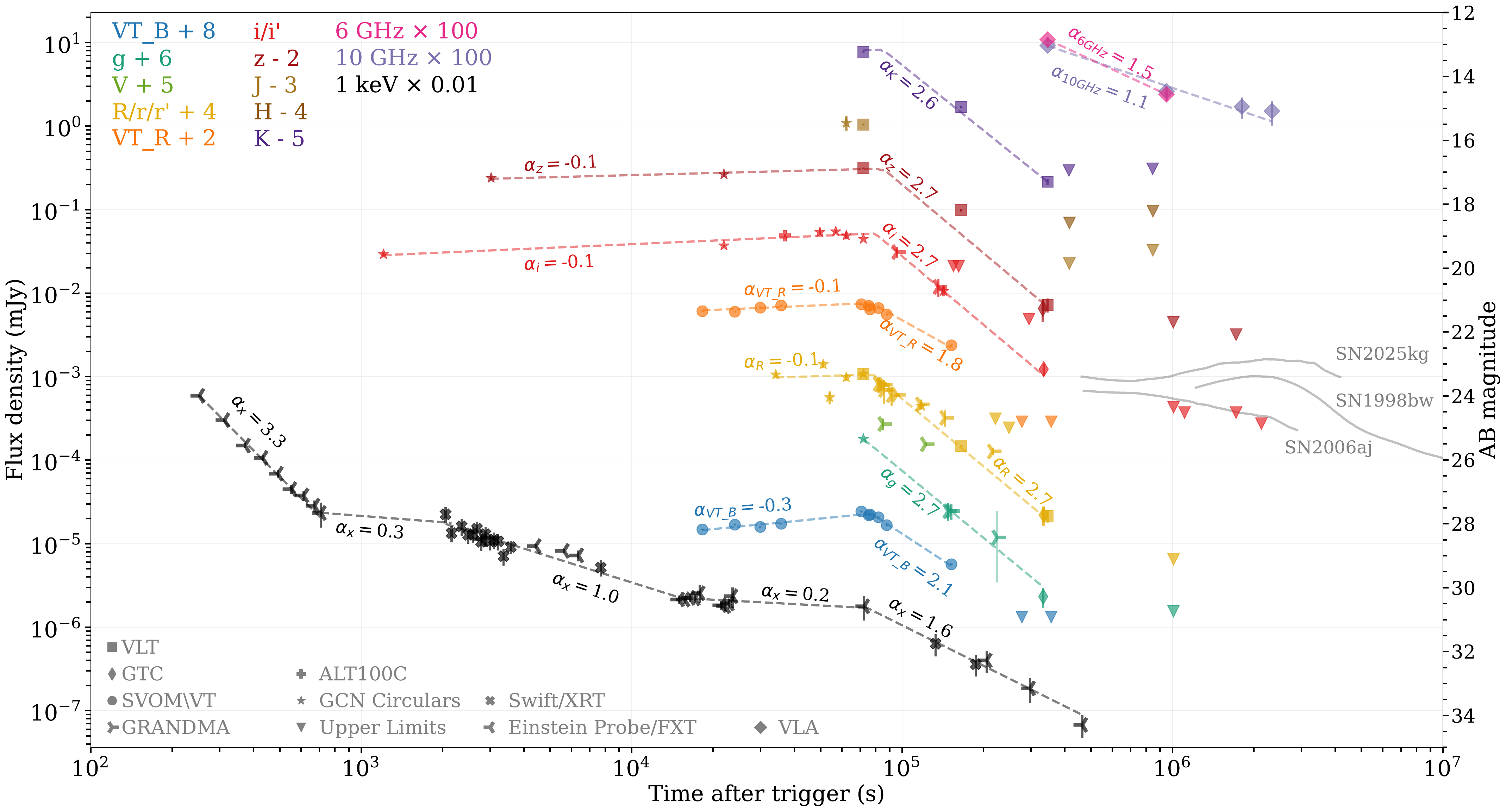}
\caption{
\textbf{Multi-band light curves of EP250704a/GRB~250704B.} 
Energy flux at different frequencies, from radio to X-rays, versus time since the GRB trigger. The temporal evolution can be described by multiple power-law segments (dashed lines). 
Error bars are at the 1$\sigma$ confidence level (c.l.). Downward triangles are upper limits at the 3$\sigma$ c.l. from VLT, SVOM/VT, GTC, OSN and GRANDMA.
Observations are compared to the optical lightcurves of known GRB-SNe: SN2025kg, SN1998bw and SN2006aj.
}
\label{fig:afterglow}
\end{figure}

\clearpage

\begin{addendum}

\item[Conflict Interests] The authors declare that they have no conflict of interest.

\item[Acknowledgments] 
This work is based on the data obtained with Einstein Probe, a space mission supported by the Strategic Priority Program on Space Science of Chinese Academy of Sciences, in collaboration with the European Space Agency, the Max-Planck-Institute for extraterrestrial Physics (Germany), and the Centre National d'E\'tudes Spatiales (France). 
We are grateful to the SVOM team for the development and operation of the SVOM mission.
The SVOM mission is supported by the Strategic Priority Research Program on Space Science of the Chinese Academy of Sciences.
This work made use of data supplied by the UK Swift Science Data Centre at the University of Leicester.
Based on observations collected at the European Organisation for Astronomical Research in the Southern Hemisphere under ESO programmes 114.27LW. 
This work was supported by the National Key Research and Development Programs of China (Grant Nos. 2022YFF0711404, 2022SKA0130102, to Bin-Bin Zhang; 2021YFA0718500, to Shao-Lin Xiong, Chen-Wei Wang; 2024YFA1611700, 2024YFA1611704, to Li-Ping Xin), the National SKA Program of China (Grant No. 2022SKA0130100, to Bin-Bin Zhang), the Strategic Priority Research Program of the Chinese Academy of Sciences (Grant Nos. XDA15360102, XDA15360300, XDB0550300, to Shao-Lin Xiong, Chen-Wei Wang; XDB0550401, to Li-Ping Xin), and the National Natural Science Foundation of China (Grant Nos. 11833003, U2038105, 12121003, 12573046, to Bin-Bin Zhang; 
12273042, 12494572, to Shao-Lin Xiong, Chen-Wei Wang, and Wen-Jun Tan; 12573049 to Hui Sun; 12494571, 12494570, 12494573, 12133003, to Li-Ping Xin). 
Bin-Bin Zhang acknowledge support by the science research grants from the China Manned Space Program (CMS-CSST-2021-B11), the Fundamental Research Funds for the Central Universities, and the Program for Innovative Talents and Entrepreneurs in Jiangsu. 
Alberto Javier Castro-Tirado acknowledges support from the Spanish Ministry project  PID2023-151905OB-I00 and Junta de Andaluc\'ia grant P20\_010168 and from the Severo Ochoa grant CEX2021-001131-S funded by MCIN/AEI/ 10.13039/501100011033. 
Alberto Javier Castro-Tirado wish to express his sincere thanks to W.H.Lee, D. Hiriart and all members of the San Pedro M\'artir Observatory in M\'exico for their assitance and operations of the 0.6m Javier Gorosabel telescope at the BOOTES-5 station. OSN and GTC optical data were collected with the 1.5m telescope at the Observatorio de Sierra Nevada (OSN) operated by the Instituto de Astrofífica de Andalucía (IAA-CSIC), and by the 10.4m Gran Telescopio Canarias (GTC) under program GTCMULTIPLE7B-25A (PI: Castro-Tirado), installed at the Spanish Observatorio del Roque de los Muchachos of the Instituto de Astrofísica de Canarias, on the island of La Palma. TRT data were based on observations made with the Thai Robotic Telescopes under program ID TRTC12B\_007, which is operated by the National Astronomical Research Institute of Thailand (Public Organization). GRANDMA thank A. Kaeouach, M. Masek, N. Sasaki, M. Lamoureux, T. Du Laz, S. Karpov, M. Freeberg, R. Hellot, K. Noysena, A. Klotz, Z. Benkhaldoun, G.M.~Hamed, M.M. Elkhateeb, A.E. Abdelaziz, M. Molham, A. Takey, Y. Rajabov, T. Kvernadze, M. Coughlin, D. Turpin, N. Guessoum, P. Hello, P-A. Duverne, S. Karpov, T. Pradier for the core team. AbAO team acknowledges Shota Rustaveli National Science Foundation of Georgia (SRNSFG). This work was supported by SRNSFG, grant FR-24-7713. NRIAG team acknowledges financial support from the Egyptian Science, Technology \& Innovation Funding Authority (STDF) under grant number 45779. This research is based on observations made with the Thai Robotic Telescope under program ID TRTC12A\_001, which is operated by the National Astronomical Research Institute of Thailand (Public Organization). This article includes observations made in the Two-meter Twin Telescope (TTT) sited at the Teide Observatory of the Instituto de Astrofísica de Canarias (IAC), that Light Bridges operates on the island of Tenerife, Canary Islands (Spain). The Observing Time Rights (DTO) used for this research were provided by Light Bridges, SL. 
GRANDMA acknowledges the french PNHE (Programme National des Hautes Energies) support. 
Eleonora Troja, Niccol\`o Passaleva, Yu-Han Yang, Rosa L. Becerra and Roberto Ricci were supported by the European Research Council through the Consolidator grant BHianca (grant agreement ID 101002761). 
Brendan O'Connor is supported by the McWilliams Postdoctoral Fellowship in the McWilliams Center for Cosmology and Astrophysics at Carnegie Mellon University. 
Daniele B. Malesani is funded by the European Union (ERC, HEAVYMETAL, 101071865). Views and opinions expressed are, however, those of the authors only and do not necessarily reflect those of the European Union or the European Research Council. Neither the European Union nor the granting authority can be held responsible for them. The Cosmic Dawn Center (DAWN) is funded by the Danish National Research Foundation under grant DNRF140.
Andrea Saccardi acknowledges support by a postdoctoral fellowship from the CNES.
Nial R. Tanvir acknowledges support from the UK STFC, grant ST/W000857/1.

\item[Author Contributions]
Bin-Bin Zhang and An Li initiated the study. 
Bin-Bin Zhang, Bing Zhang coordinated the scientific investigations of the event.
Yi-Han Iris Yin analyzed the high energy prompt emission data. 
Chen-Wei Wang, Shao-Lin Xiong, Wen-Jun Tan and Chao Zheng contributed to the data analysis of SVOM/GRM and Insight/HXMT.
Jia-Xin Cao and Xin-Lei Chen are the GRM burst advocates on July 4 2025.
Yong-Wei Dong, Cheng-Kui Li, Xiao-Bo Li, Wen-Jun Tan, Chen-Wei Wang, Shao-Lin Xiong, Shu-Xu Yi, Shuang-Nan Zhang, Chao Zheng and Shi-Jie Zheng contributed to the development and operation of SVOM/GRM and Insight/HXMT. 
Li-Ping Xin and Hua-Li Li processed the SVOM/VT data. 
An Li, Ye-Hao Cheng, Chang Zhou, Guo-Ying Zhao, Yi-Jia Zhang and Wen-Da Zhang are the WXT transient advocates on July 4 2025 and contributed to the discovery and preliminary analysis of the event.
Yuan Liu, Hui Sun and Qin-Yu Wu contributed to the WXT data processing. 
Zhi-Xing Ling, Chen Zhang and Weimin Yuan contributed to the development of the WXT instrument. 
Chen Zhang, Zhi-Xing Ling, Hua-Qing Cheng and Yuan Liu contributed to the calibration of WXT data.
Yuan Liu, Hua-Qing Cheng, Hai-Wu Pan, Dong-Yue Li, Jing-Wei Hu, He-Yang Liu and Hui Sun contributed to the development of WXT data analysis software.
Yong Chen contributed to the development of the FXT instrument. Shu-Mei Jia contributed to the development of FXT data analysis software.
Yu-Lei Qiu, Jing Wang, Xu-Hui Han, Jian-Yan Wei and Bertrand Cordier contributed to the operation of VT. 
Jie An, Dong Xu, Valerio D'Elia, Massimiliano De Pasquale, Rob A. J. Eyles-Ferris, Dieter Hartmann, Andrew J. Levan, Daniele B. Malesani, Dani\"elle L.~A. Pieterse, Andrea Saccardi, Nial R. Tanvir, Susanna D. Vergani contributed to the discovery and early characterization of the afterglow and provided feedback on the paper.
Maria Gritsevich, Shashi Bhushan Pandey, Rub\'en S\'anchez-Ram\'irez and Alberto Javier Castro-Tirado prepared GTC proposal and processed its data, Rub\'en S\'anchez-Ram\'irez also analysis data from OSN. 
Ignacio Perez-Garcia analysis data from BOOTES-5/JGT. 
Sarah Antier, Eslam Elhosseiny, Nino Kochiashvili, Marion Pillas, Tillayev Yusufjon, Manasanun Tanasan and Xiao-Feng Wang contributed to the operation and data analysis of GRANDMA.
Eleonora Troja, Rosa L. Becerra, Niccol\`o Passaleva and Yu-Han Yang led the VLT campaign and were responsible for data reduction and calibration. 
Niccol\`o Passaleva and Yu-Han Yang were responsible for the X-shooter spectral analysis. 
Brendan O'Connor and Eleonora Troja led the VLA campaign, Roberto Ricci was responsible for the radio data reduction and calibration. 
Hou-Jun L\"{u}, He Gao provided idea about the EE comparison, and An Li search the Swift sample and made comparison. 
Yi-Xuan Shao and You-Dong Hu made analysis to the late time optical observations. 
Niccol\`o Passaleva, An Li, Jie An, Dong Xu and He Gao made afterglow fitting and modeling process. 
An Li, Yi-Han Iris Yin, Jun Yang, Wei-Hua Li, Eleonora Troja, Bin-Bin Zhang, and Bing Zhang further investigated the physical model, completed the schematic diagram, and contributed to the discussion of the physical implications. 
Bin-Bin Zhang, Eleonora Troja, Yi-Han Iris Yin, An Li and Bing Zhang wrote the manuscript, with contributions from all authors.

\end{addendum}

\clearpage

\clearpage
\noindent{\bf References}
\bigskip
\bigskip

\bibliographystyle{SciBull}
\bibliography{main}

\clearpage

\section*{\centering \large Supplementary material}

\subsection{Observations and data reduction.\\}

Logs of the observations used in this work are presented in Table~\ref{tab:optical_obs}.

\noindent\textbf{EP-WXT.} 
The Einstein Probe (EP), led by the Chinese Academy of Sciences (CAS) in collaboration with the European Space Agency (ESA) and the Max Planck Institute for extraterrestrial Physics (MPE), Germany, is a mission focused on the time-domain high-energy astrophysics. Among its two main payloads is Wide-field X-ray Telescope (WXT), which utilizes innovative lobster-eye micro-pore optics (MPO), enabling a large instantaneous field of view (FoV) of 3,600 square degrees and a sensitivity of $\sim 2.6\times 10^{-11} \rm erg/cm^{2}/s$ in 0.5–4 keV with 1 ks exposure. EP250704a was detected by the Wide-field X-ray Telescope (WXT) onboard the Einstein Probe (EP) during a monitoring target observation of 2RXS J210955.2+480922 from UTC 2025-07-04T07:49:30 to 2025-07-04T08:16:52 (ObsID: 11916650016). This source triggered the on-board processing unit 25 s after the Space Variable Objects Monitor (SVOM)/Gamma-ray Burst Monitor (GRM) trigger, and the autonomous EP-Follow-up X-ray Telescope (FXT) follow-up was preformed 221s after the WXT trigger. We recovered the WXT continued observation data during the slew phase. The WXT observation continued for $\sim 800$ s after the trigger, and the burst emission before 334s was detected with a signal-to-noise ratio higher than $3\sigma$. The X-ray events data were selected and calibrated using the data reduction software and calibration database (CALDB) designed for WXT (Liu et al. in prep.). The CALDB is generated based on the results of the on-ground calibration experiments \cite{2025arXiv250518939C}, the procedure of which has been applied to a prototype of WXT instrument. The image in the 0.5–4 keV range was extracted from the cleaned events (Fig. \ref{fig:detection}). The light curve and the spectrum of the source and background in a given time interval were extracted from a circle with a radius of 9 arcmin and an annulus with the radii of 18 arcmin and 36 arcmin, respectively. 

\noindent\textbf{SVOM-GRM.} The Space-based multi-band astronomical Variable Objects Monitor (SVOM) mission is a Franco-Chinese mission dedicated to the study of GRB. It was launched on June 22, 2024. The Gamma-Ray Monitor (GRM) is one of its main instruments designed to measure the spectrum of high-energy bursts. GRM comprises three scintillator-based gamma-ray detectors (GRDs) operating in the 15 keV--5 MeV energy range onboard the Sino-French satellite SVOM \cite{GRM_intro,GRM_trigger}. On 2025-07-04T08:16:27.100 UTC (denoted as $T_0$), all the three GRDs were triggered by a transient event with a signal-to-noise ratio of 29.6 in the 0.1-second time window. The real-time alert data of this transient was downlinked to the ground through the VHF system with low latency, and then identified as a short burst GRB~250704B by the standard analysis process \cite{BA_pipeline}. Based on the preliminary result of VHF data, GRB~250704B was reported to the community by SVOM/GRM firstly \cite{2025GCN.40940....1S}. The event-by-event data, which is well calibrated (in preparation) in 15-5000 keV, is used for detailed temporal and spectral analysis. 

\noindent\textbf{Insight-HXMT.} As China's first X-ray astronomy satellite, Insight-HXMT consists of three groups of collimated telescopes, namely the high energy X-ray telescope (HE), the medium energy X-ray telescope (ME) and the low energy X-ray telescope (LE) \cite{HXMT_zhang_2018,HXMT_zhang_2020,HXMT_Li_2020}. GRB~250704B was detected by the CsI detectors of HE at $T_0$, which have a large effective area for gamma-ray photons penetrating the satellite platform \cite{2025GCN.40978....1W}. The energy range of CsI data used for spectral analysis is 120--600\,keV, while for temporal analysis is 60--900\,keV \cite{HE_calibration,HE_calibration_updated}. The data reduction follows the standard analysis process \cite{BA_pipeline}.

\noindent\textbf{EP-FXT.} The Follow-up X-ray Telescope (FXT) is one of the main payloads of EP in the 0.3–10 keV band, which offers three readout modes, i.e., Full Frame mode (FF), Partial Window mode (PW) and Timing mode (TM). The FXT is designed for quick follow-up observations and more precise positioning of transients, and for observations of targets of opportunity. EP-FXT carried out 6 follow-up observations of EP250704a/GRB~250704B, spanning from 221 seconds to approximately 5.4 days after the burst. The cumulative observation time exceeded $5\times 10^4$ s. During these observations, FXT-A observed in PW mode and FXT-B observed in FF mode in autonomous follow-up observation, both FXT-A and FXT-B observed in FF mode in later follow-up observations. All observations were conducted with a thin filter. The FXT data were processed using the FXT Data Analysis Software (FXTDAS v1.20), utilizing the latest FXT calibration database (CALDB v1.10). Source photons were extracted from a circle with a radius of $40$ arcsec, while background photons were extracted from an annulus with the radii of 60 arcsec and 180 arcsec. FXT-B suffered from pile-up in FF mode for the first $\sim$ 600 s in the autonomous follow-up observation, during which time interval we only utlized the data from FXT-A in PW mode.

\noindent\textbf{Swift-XRT.} Swift-XRT has performed 4 follow up observations to EP250704a/GRB~250704B ranging from 2025-07-04T08:46:08 UTC to 2025-07-08T00:09:56 UTC with an exposure time of 11ks. All the observations were conducted in PC mode. The spectra and flux curves from XRT were selected by a circle source region with 30 pixel radius and an annuli background region with 80 to 120 pixel radius, and calibrated using the data reduction software in HEASoft6.35.5 \cite{2012ApJS..203....2D} package and the latest calibration database (CALDB) designed for XRT. To keep consistence with FXT observation, the flux curve were extracted within 0.5 to 10.0 keV energy range by \texttt{xrtgrblc} in HEASoft tools, same as the online tools result \cite{2007A&A...469..379E,2009MNRAS.397.1177E}.

\noindent\textbf{Optical and Near-Infrared imaging} 
We performed follow-up photometric observations with multiple instruments, including: ESO VLT UT1 (Antu) equipped with the FORS2 camera \cite{1998Msngr..94....1A} under program 114.27LW (PI: E. Troja), ESO VLT UT4 (Yepun) equipped with the near-infrared HAWK-I imager under program 114.27LW (PI: E. Troja), 100-cm C telescope (ALT100C) of the JinShan project, located in Altay, Xinjiang, China, the 0.7-m telescope of the Thai Robotic Telescope network (TRT), located at Fresno, California, U.S.A (SRO), the 0.6-m Javier Gorosabel Telescope at the BOOTES-5 astronomical station in San Pedro M\'artir Observatory (M\'exico), Visible telescope (VT) on board SVOM \cite{2016arXiv161006892W}, 1.5 m telescope of Sierra Nevada Observatory (OSN, Granada, southern Spain,\url{http://www.osn.iaa.es/}), the 10.4m Gran Telescopio CANARIAS (GTC) equipped with the Optical System for Imaging and low-Intermediate-Resolution Integrated Spectroscopy (OSIRIS) instrument for optical bands and the Espectrógrafo Multiobjeto Infra-Rojo (EMIR) instrument for NIR bands, and the Global Rapid Advanced Network Devoted to the Multi-messenger Addict (GRANDMA \cite{GRANDMA_2020}) contains kilonova-catcher telescopes (\url{http://kilonovacatcher.in2p3.fr/}), Ulugh Beg Astronomical Institute telescopes (UBAI NT-60 and UBAI AZT-22 at Maidanak Observatory), Kottamia Astronomical Observatory (KAO) telescopes, the ASO telescope at Oukaimeden Observatory and the Abastumani Astrophysical Observatory Telescope (AbAO-T150). Photometric results were derived using aperture photometry through standard procedures in the IRAF software. For GRANDMA results they were derived using the STDPipe pipeline and its web interface, to perform force photometry and template subtraction on a set of images obtained from different telescopes \cite{Karpov_2025}.

Photometric calibration in the $griz$ bands was performed as follows. For VLT, TRT, and ALT100C observations, we used the Pan-STARRS DR2 catalog \cite{2018AAS...23143601F}. The OSN observations were calibrated against field stars listed in the SDSS DR8 catalog \cite{SDSSdr8}. For instruments within the GRANDMA network, we used the Gaia DR3 catalog \cite{Gaia2023} for the $R$, $V$, and $I$ bands, and the Pan-STARRS DR1 catalog \cite{chambers2019panstarrs1surveys} for the $g$, $r$, and $i$ filters. SVOM/VT, VLT observations were directly calibrated in the AB magnitude system. For $BVRI$ band photometry only, we adopted the transformation relations from \url{https://www.sdss4.org/dr12/algorithms/sdssUBVRITransform/\#Lupton2005}. For the near-infrared(NIR) band they were corrected by 2MASS Point Source Catalog \cite{2mass+2003}. The obtained photometric magnitudes in AB system and corrected for Galactic dust extinction, following the estimates of ref \cite{Gal_ext}, are listed in Table \ref{tab:optical_obs}. 

For GTC's observations, due to high contamination from the wings of the point-spread function (PSF) of a nearby bright star, which affected background estimation at the position of GRB~250704B, PSF subtraction was performed prior to photometric analysis. We constructed an empirical PSF model using the \texttt{photutils.psf} routine, based on a set of unsaturated and isolated field stars. This model was used to fit and subtract the bright star near the GRB. A relationship between instrumental and absolute magnitudes was established by fitting the background-subtracted flux of these reference stars. This relation was then used to convert the background-subtracted flux of the target into absolute magnitudes. Finally, we applied the Vega-to-AB magnitude transformation \cite{Blanton+2005AJ} to report all results in the AB system, as provided in Table \ref{tab:optical_obs}.

\noindent\textbf{Radio.} 
We observed the source with the Karl G. Jansky Very Large Array (VLA) \cite{1998AJ....115.1693C} in four X-band epochs at 10 GHz with 4 GHz bandwidth, under program 25A-309 (PI: Troja) and two C-band epochs at 6 GHz with 4 GHz bandwidth, under program SN078192 (PI: O'Connor).
All data were obtained in the array configuration C, using 3C286 as the primary flux calibrator and J1950+0807 as the phase calibrator. Raw data were retrieved from the NRAO archive and processed with CASA v6.6.1 \cite{2022PASP..134k4501C} using the standard VLA pipeline for flagging, bandpass, gain, and flux calibration. Imaging was performed with tclean adopting a Briggs weighting of 0.5 and H\"ogbom deconvolution. The rms noise was measured in a source-free region, and a $5\%$ calibration uncertainty was added in quadrature to the statistical error. Flux densities were determined from two-dimensional Gaussian fits, with the peak flux taken as the best estimate for point-like sources.

\subsection{Spectral Analysis.\\}

\noindent\textbf{Independent fit with gamma-ray spectra.} 
We conducted both time-integrated and time-resolved spectral analyses using gamma-ray data from SVOM/GRM (15--5000 keV) and Insight-HXMT (120--600 keV). The time intervals were determined according to individual pulses. The spectral fittings were carried out with the Python package, bayspec, following the methodology described in ref \cite{Yin_2025}. The fittings were evaluated using PGSTAT \cite{1996ASPC..101...17A}, which is well suited for gamma-ray observations where the source spectrum follows Poisson statistics while the background spectrum is characterized by Gaussian errors. The spectral models adopted in the spectral fittings are summarized below:
\begin{itemize}
\item The power law (PL):
\begin{equation}
N(E)=A\Big(\frac{E}{100\,{\rm keV}}\Big)^{\alpha},
\end{equation}
where $A$ is the normalization parameter in units of ${\rm photons~cm^{-2}~s^{-1}~keV^{-1}}$, and $\alpha$ is the spectral index at low energies.
\item The cutoff power law (CPL):
\begin{equation}
N(E)=A\Big(\frac{E}{100\,{\rm keV}}\Big)^{\alpha}{\rm exp}\Big(-\frac{E}{E_{\rm c}}\Big),
\end{equation}
where $A$ is the normalization parameter, $\alpha$ is the spectral index at low energies, and $E_{\rm c}$ is the cutoff energy in units of keV. The peak energy $E_{\rm p}=(2+\alpha)E_{\rm c}$.
\item The smoothly broken power law with high-energy exponential cutoff (CSBPL):
\begin{equation}
N(E)=AE_{\rm b}^{\alpha_1}\Big[\Big(\frac{E}{E_{\rm b}}\Big)^{-\alpha_1 n}+\Big(\frac{E}{E_{\rm b}}\Big)^{-\alpha_2 n}\Big]^{-\frac{1}{n}}{\rm exp}\Big(-\frac{E}{E_{\rm c}}\Big),
\end{equation}
where $A$ is the normalization parameter,
$\alpha_1$ and $\alpha_2$ are the spectral indices of the two smoothly-connected law segments at the break energy $E_{\rm b}$. The smoothness parameter $n$ is fixed to 5.38 \cite{2018A&A...613A..16R}. $E_{\rm p}=(2+\alpha_2)E_{\rm c}$. 
\end{itemize}

We fitted the gamma-ray spectra with spectral models featured with high-energy exponential cutoffs, CPL and CSBPL, considering the limited instrumental coverage at the high-energy end. The models are compared based on the Bayesian Information Criterion (BIC) \cite{bic_ref}, and both provided comparable fits ($\Delta \rm{BIC} < 5$). The best-fit parameters and fitting statistics for each interval are listed in Table \ref{tab:spike_spec_fit}, and the corresponding spectral parameter evolutions are presented in Fig.~\ref{fig:HE_spec_evo}.

\noindent\textbf{Independent fit with X-ray spectra.} We fitted the EP/WXT spectra with the absorbed PL model, \textit{tbabs*ztbabs*pl}, adopting a Galactic absorption column density of $N_{\rm H,gal}=8.94\times10^{20}~\rm{cm^{-2}}$ and an intrinsic absorption column density of $N_{\rm H,int}=6.52\times10^{21}~\rm{cm^{-2}}$, as determined from the EP/FXT spectral fittings with adequate statistical constraints (Methods). Time intervals are determined via adaptive rebinning of the light curve, requiring $\ge 40$ background-subtracted net counts in each bin, which are subsequently used for spectral fitting. Owing to the limited photon statistics, the WXT spectrum of the sharp spike does not provide meaningful constraints on the spectral shape and was thus analyzed through joint spectral fitting. The best-fit parameters and corresponding statistics for each time slice are listed in Table \ref{tab:ep_spec_evo}. Based on the spectral fittings, we derived the spectral evolution, shown in Fig.~\ref{fig:detection}. Generally, the spectral index follows an intensity-tracking pattern \cite{1983Natur.306..451G} up to $\sim T_0 + 200$ s, with hard values of $\alpha\sim -1$ in the initial spike and softer values of $\alpha\sim -2$ in the long bump. A more resolved spectral evolution study was performed with EP/FXT data from $T_0 + 222$ s onward, where the photon statistics are substantially improved.

\noindent\textbf{Joint fit with EP/WXT, SVOM/GRM and Insight-HXMT.} The joint fit of the second pulse, spanning $T_0+[0.1,0.7]$ s, was carried out with both an absorbed CPL model \textit{tbabs*ztbabs*cpl} and CSBPL model \textit{tbabs*ztbabs*csbpl}. 
The parameters of the absorption models for the WXT spectrum were fixed to the values adopted from the independent fits of EP/FXT spectra, as detailed in the Methods below.
The best-fit parameters and corresponding statistics are listed in Table~\ref{tab:spike_spec_fit}. The CPL parameters obtained from the independent fit to the gamma-ray spectrum in $T_0+[0.1, 0.4]$ s agree well with those from the joint fit, indicating that the spectral shape is primarily constrained by the gamma-ray data, which have higher photon statistics. Additionally, both CPL and CSBPL provide comparably good fits, with a BIC difference of less than 5. However, the CPL fitting shows residuals exceeding $2~\sigma$ in the X-ray band, whereas the CSBPL fitting keeps all deviations within $2~\sigma$. Consequently, the joint analysis favors the CSBPL model for the spectrum of the second pulse.

\noindent\textbf{Spectral hardening transition in the long-duration X-ray detection with EP/FXT.} We conducted time-resolved spectral fitting of the EP/FXT data using the Python package bayspec (\url{https://github.com/jyangch/bayspec}), an upgraded version of MySpecFit, following the methodology outlined in ref \cite{Yin_2025}. The time slices were divided by requiring a minimum of 200 accumulated photon counts to ensure adequate statistical significance in each time-resolved spectrum. Considering the brightness-induced pileup affecting FXT-B in full-frame mode, only FXT-A data were used for the spectral analysis within $T_0$+[222, 717] s. For epochs later than $T_0$+3460 s, joint spectral fittings were performed using spectra from both FXT-A and FXT-B. The goodness of fit was assessed by the reduced statistic STAT/dof, as described in ref \cite{2024ApJ...975L..27Y}, while model comparison employed the Bayesian information criterion (BIC) \cite{bic_ref}. The best-fit parameters and fitting statistics for each interval are listed in Table \ref{tab:ep_spec_evo}, and the corresponding spectral index evolution is presented in Fig.~\ref{fig:fxt_spec_evo}.

The time-resolved spectral analysis reveals clear variable absorption across $T_0$+375.5 s. Specifically, when fitting with the \textit{tbabs*ztbabs*pl} model, the intrinsic absorption column density, $N_{\rm H,int}$, was constrained to $6.52_{-1.07}^{+1.17}\times10^{21}~\rm{cm^{-2}}$ during $T_0$+[222.0, 375.5] s, but declined to an unconstrained, negligible value afterward. Consequently, we adopted the model \textit{tbabs*ztbabs*pl} with fixed intrinsic absorption column density $N_{\rm H,int}=6.52\times10^{21}~\rm{cm^{-2}}$ and galactic absorption column density $N_{\rm H,gal}=8.94\times10^{20}~\rm{cm^{-2}}$ for spectra prior to $T_0$+375.5 s, and \textit{tbabs*pl} with fixed $N_{\rm H,gal}=8.94\times10^{20}~\rm{cm^{-2}}$ for subsequent intervals.

The derived spectral indices, $\alpha_{\rm X}$, exhibit a distinct soft--hard--flat spectral evolution pattern, which indicates the transition from the fading prompt emission to the emerging afterglow at around $T_0$+350 s \cite{Yin_2025}. The afterglow dominates the emission at later times from the observation starting at $T_0$+3460 s, displaying a characteristic normal temporal decay expected for this phase.

\noindent\textbf{VLT/X-shooter Optical Spectroscopy.} EP250704a lies 13 arcsec away from a bright ($r\sim$18 AB mag) nearby ($z=0.1062$) galaxy, SDSS J200329.53+120109.9, which was initially considered as a possible host given its low chance-coincidence probability ($P_{\rm cc}\!\sim2$\%) \cite{2002AJ....123.1111B}. To establish the source distance scale and its possible relation to the SDSS galaxy, spectroscopic observations were conducted with the X-shooter spectrograph \cite{2011A&A...536A.105V} mounted on UT3 (Melipal) of ESO's Very Large Telescope (VLT). 

Here we utilize data from Programme 114.27LW (PI: E. Troja) and present the corresponding analysis and results. 
The observations consisted of 4$\times$1200 s exposures taken at an average airmass of 1.26 and seeing $\sim$0.83 arcsec. A slit width of 1.0 arcsec was used in the UVB arm and 0.9 arcsec in the other arms.
Individual exposures were reduced using the official ESO Reflex pipeline \cite{2013A&A...559A..96F} and then median combined into a single stacked spectrum. Superposed on a bright continuum, we identify multiple absorption lines due to Fe II, Mg I,  and Mg II at a common redshift of $z$ = 0.66102 $\pm$ 0.00011 (Fig.~\ref{fig:spec_xshooter}), thereby ruling out the association with the nearby galaxy. This is consistent with the preliminary estimate reported in \cite{2025GCN.40966....1A}.

\subsection{Interpretation of  spike consistency and divergence in Episodes I and II.\\}

\noindent\textbf{Gamma-ray first pulse without X-ray counterpart.} Based on the independent and joint spectral fitting results of the first and second pulses of GRB 250704B, we constructed the spectral energy distributions (SEDs), illustrated in Fig. \ref{fig:spike_sed}. The single pulse detected in EP250704a is both temporally and spectrally consistent with the second pulse in GRB 250704B (Fig.~\ref{fig:consistency}), whereas no X-ray counterpart is associated with the first pulse. 
Assuming that the prompt X-ray and gamma-ray emissions belong to the same single synchrotron radiation component, the spectral behavior between the X-ray pulse and the second gamma-ray pulse could be described by a synchrotron-like CSBPL model. This model also fits well with the spectrum of the first gamma-ray pulse, yielding a low-energy spectral index $\alpha \sim -0.97_{-0.24}^{+0.56}$, a break energy $E_{\rm break} \sim 54_{-38}^{+52}$ keV, and a peak energy $E_{\rm p} \sim 318_{-79}^{+76}$ keV. The spectral shape is broadly consistent with the synchrotron emission scenario in the regime $\nu_a < \nu_m < \nu_p$, where the expected spectral index between $\nu_a$ and $\nu_m$ is $\sim -2/3$, consistent within uncertainties with the measured $\alpha$. From the X-ray non-detection at the time of gamma-ray first pulse, we derived upper limits that fall below the low energy extrapolation of the best-fit absorbed CSBPL model, which indicates a spectral break between the optical and X-ray bands. To account for an additional spectral break below $\nu_m$, it is natural to invoke synchrotron self-absorption \cite{2009MNRAS.398.1936S} as a plausible explanation, in which synchrotron photons are re-absorbed by the same population of relativistic electrons that produced them.
By applying a broadband synchrotron spectrum for first pulse of GRB~250704B, with a low-energy spectral slope of $f_{\nu}\propto\nu^2$, we obtained a lower limit for the break energy of $\nu_a \sim 6$ keV, resulting a theoretical flux just below the detection upper limits.

\noindent\textbf{Consistency of the second pulse in X-rays and gamma-rays.} The second pulse of the GRB~250704B spike, spanning $T_0+[0.1,0.35]$ s, temporally overlaps with the single spike detected in EP250704a within $T_0+[0.2,0.7]$ s. We modeled the second pulse with a Gaussian profile across different gamma-ray energy bands, revealing an anti-correlation between the full width at half maximum (FWHM) and energy. Confined by the limited photons in X-ray spike, we only described the pulse with a Gaussian model using morphologically reasonable parameters (Fig. \ref{fig:consistency}). The resulting X-ray pulse width lies on the low-energy extrapolation of the second gamma-ray pulse, supporting their association. Spectrally, a joint fit of the WXT, GRM, and HXMT data over $T_0+[0.1, 0.7]$ s with an absorbed CSBPL model yields a good fit, consistent with a single emission component spanning from X-rays to gamma-rays. Notably, there is a peak time difference between the X-ray and gamma-ray pulses, a feature also seen in other associated EFXTs/GRBs \cite{2024ApJ...975L..27Y, Yin_2025}. 
While such an offset is potentially intriguing if it significantly deviates from the energy-dependent spectral lag trend \cite{2006ApJ...653L..81L, 2016ApJ...825...97U}, the current data do not allow a robust assessment of such a deviation. Within the uncertainties, the peak time between the X-ray and gamma-ray pulses remains consistent with an extrapolation of the energy-dependent peak time difference relation toward lower energies in logarithmic space (Fig. \ref{fig:consistency}b). This is broadly consistent with a common origin, as expected for a single emission component spanning from X-rays to gamma-rays. Future observations with improved photon statistics will be essential to better characterize and test any potential intrinsic offset between the X-ray and gamma-ray peaks.

\noindent\textbf{X-ray tail lacking a gamma-ray counterpart.} Following the initial spike of Episode I, a decaying X-ray tail at $\sim T_0+[1.5, 3.5]$ s shows a consistently hard spectral index but without corresponding gamma-ray emission. This behavior aligns with the gradual flux decline from the second pulse of Episode I to Episode II. Assuming the same spectral shape as the second pulse of Episode I, the 15--5000 keV flux ($\sim 1.44\times 10^{-6}~\rm{erg~cm^{-2}~s^{-1}}$) is slightly below the SVOM/GRM sensitivity of $\sim 2.37\times 10^{-6}~\rm{erg~cm^{-2}~s^{-1}}$.

\subsection{Engine-powered internal origin of Episode III.\\}

\noindent\textbf{Comparison of relative variability timescale and flux with theoretical limits.} The variability of Episode III is determined by employing the Bayesian method \cite{Scargle_2013} to recognize the minimum Bayesian block. We obtain a relative variability timescale of $\Delta t/t \sim 9.5\times 10^{-2}$ and a relative flux amplitude of $\Delta F_{\nu}/F_{\nu} \sim 3.8$. We compare these values with theoretical limits of variability from the afterglow \cite{2005ApJ...631..429I}. The first is the variability caused by density fluctuations in the circumburst medium. As the blast wave propagates into regions of different density, fluctuations in the afterglow flux can be produced. However, the amplitude of such variability is limited by the constraints derived by \cite{2005ApJ...631..429I} for both on-axis and off-axis cases. The second scenario is the patchy-shell model, in which the emitting surface has an intrinsic angular structure. In this case, the observed flux varies as different angular regions enter the visible area. Since the visible region expands on a timescale comparable to $t$, the variability timescale is expected to satisfy $\Delta t \gtrsim t$. The third scenario is the refreshed shocks, where slower ejecta launched by the central engine catch up with the decelerating leading shell and inject additional energy into the external shock. Such energy injection produces variability with a timescale of $\Delta t \gtrsim t/4$. We find that EP250704a lies in a region that violates all of these limits, as presented in Fig. \ref{fig:variability_plane}. This result supports the presence of a still active central engine during the minutes-long soft X-ray extended emission (Episode III).

\noindent\textbf{Analysis of the steep decay.} The steep decay phase of Episode III has a temporal index of $\sim -2.9_{-0.2}^{+0.2}$ (see below), which is significantly steeper than the canonical fallback accretion slope of $\sim -5/3$ \cite{1988Natur.333..523R, 2007MNRAS.376L..48R}, indicating that the emission of Episode III is inconsistent with being driven solely by fallback accretion onto a newly formed black hole, although a black hole central engine cannot be completely ruled out \cite{2015ApJ...804L..16K, 2023ApJ...958L..33G}. We further tested the dependence of the inferred decay index on the choice of the zero time. Although Episode III begins at $\sim T_0+20$ s, we conservatively adopted a larger shifted zero time of $\sim T_0+50$ s, obtaining a lower limit on decay index of $2.6_{-0.2}^{+0.1}$. Although reduced, this slope remains significantly steeper than the canonical fallback accretion slope and the normal external shock afterglow decay, but is consistent with an internal plateau.

Combining both the temporal and spectral slopes, we further examine the steep decay phase against the expectations of the curvature effect \cite{2000ApJ...541L..51K, 2004ApJ...614..284D}, which predicts a characteristic relation between the temporal decay index $\alpha$ and the spectral index $\beta$ following the abrupt cessation of the prompt emission region \cite{2006ApJ...642..354Z, 2006ApJ...646..351L, 2009ApJ...690L..10Z} We converted the spectral fitting index $\alpha_{\rm X}$ in Table \ref{tab:ep_spec_evo} into the spectral index $\beta$ in the convention $F_\nu \propto t^{-\alpha} \nu^{-\beta}$ \cite{2000ApJ...541L..51K, 2004ApJ...614..284D, 2015ApJ...808...33U}, and then derived the curvature effect predicted temporal slope from the relation $\alpha=2+\beta$. As shown in Fig. \ref{fig:curvature}, the observed temporal decay is shallower than the prediction of the curvature effect before the spectral hardening at the end of the decay. This indicates that the emission cannot be explained solely by high-latitude emission following the prompt phase, and instead suggests that the central engine has not yet shut down during this phase.\\
We also attempted to model the steep decay within a standard external forward-shock framework, including the case of a jet break, but find that reproducing $t_{\rm break}\sim 318$ s requires an extreme jet opening angle of $\sim$0.003 rad, smaller than any short GRB reported to date, disfavoring this scenario.

\noindent\textbf{Magnetar parameters estimation and NS equation-of-state constraints.} We fit the X-ray light curve of Episode III over the interval $T_0$ + 5 s to 1000 s with two broken power-law models, accounting for the underlying plateau followed by the steep decay and the flare (Figure \ref{fig:comparison}). The fitting results show that the underlying component is consistent with a nearly flat plateau followed by a steep decay with index of $-2.9_{-0.2}^{+0.2}$, supporting the interpretation of an internal plateau. From the best-fit parameters, we derive the plateau X-ray luminosity of $L_{\rm X}=2.1_{-0.6}^{+1.2}\times10^{47}~{\rm erg~s^{-1}}$ and the spin-down timescale of $\tau=318.5_{-25.0}^{+25.5}$ s. We further convert the X-ray luminosity $L_X$ to the rest-frame bolometric luminosity in the 1--10000 keV band, obtaining $L_0 \sim 3.1\times10^{47}$ erg s$^{-1}$, by applying the $k$-correction based on the spectrum of the plateau phase.

Assuming that a millisecond magnetar loses its rotational energy predominantly through magnetic dipole spin down radiation \cite{2008MNRAS.385.1455M}, the initial spin period $P_0$ and the surface magnetic field strength $B_{\rm p}$ can be inferred from the characteristic plateau luminosity $L_0$ and the spin-down timescale $\tau$ through \cite{2015ApJ...805...89L}

\begin{equation}
B_{\rm p, 15} = 2.05~{\rm G}~(I_{45}R_{6}^{-3}L_{0, 49}^{-1/2}\tau_{3}^{-1}),
\end{equation}
\begin{equation}
P_{0, -3} = 1.42~{\rm s}~(I_{45}^{1/2}L_{0, 49}^{-1/2}\tau_{3}^{-1/2}),
\end{equation}
where $I$ is the moment of inertia and $R$ is the radius of the neutron star. Here we tentatively adopt a canonical magnetar mass of $1.4~M_\odot$ and radius of $10^{6}$ cm. The initial spin period $P_0$ and the surface magnetic field strength $B_{\rm p}$ of EP250704a magnetar are presented in Fig. \ref{fig:magnetar_params}. They are consistent with those obtained from other short GRB magnetar central engines \cite{2013MNRAS.430.1061R}.

The neutron star Equation of State (EOS) can be constrained from the inferred $B_{\rm p}$, $P_0$, and collapsed time. For a given EOS, the collapse time is expressed as a function of the protomagnetar mass\cite{2014PhRvD..89d7302L, 2015ApJ...805...89L} as

\begin{equation}
t_{\rm col}=\frac{3c^3I}{4\pi^2B_{p}^2R^6}[(\frac{M_{\rm p}-M_{\rm TOV}}{\alpha M_{\rm TOV}})^{2/\beta}-P_0^2].
\end{equation}
where $\alpha$ and $\beta$ and $M_{\rm TOV}$ depend on the EOS and are taken from \cite{2014MNRAS.441.2433R, 2014PhRvD..89d7302L}. For EP250704a, we adopt the inferred $B_{\rm p}$ and $P_0$ using the canonical magnetar mass and radius, and plotted the collapse time-protomagnetar mass relations for different EOS models in Fig. \ref{fig:magnetar_params}. Assuming a collapse time of $t_{\rm col} \sim 2$ ks, the corresponding protomagnetar mass for the GM1 EOS falls within the 68\% confidence interval of the protomagnetar mass distribution derived from the Galactic NS-NS binary systems \cite{2013arXiv1309.6635K}. This indicates that inferred magnetar parameters are more consistent with the GM1 EOS.

As suggested by the EOS model constraints, we also use the GM1 EOS values, i.e., $M=2.37~M_\odot$ and $R=1.205\times10^{6}$ cm \cite{2014PhRvD..89d7302L, 2015ApJ...805...89L}, to further infer the surface magnetic field strength and initial spin period. The resulting parameters are in agreement with those of previously studied magnetar central engines, as shown in Fig. \ref{fig:magnetar_params}.

\subsection{Statistical analysis.\\}

\noindent\textbf{Rest-frame peak energy ($E_{\text{peak},z}$)–isotropic gamma-ray energy ($E_{\gamma,\text{iso}}$)}: 
The $E_{\text{peak},z}$-$E_{\gamma,\text{iso}}$ relation, also named as Amati relation \cite{2002A&A...390...81A}, is assumed to follow a linear form:\[\log E_{\text{peak},z} = b + k \log E_{\gamma,\text{iso}}.\]To derive the best-fitting values and their uncertainties based on previous GRB samples with known redshift \cite{2020MNRAS.492.1919M}, we use the MCMC method \cite{2013PASP..125..306F}, which provides Bayesian posterior distributions for the model parameters. The results, with 3$\sigma$ uncertainties, are as follows: for Type I GRBs, $k = 0.36^{+0.04}_{-0.05}$, $b = -15.61^{+2.51}_{-2.14}$; for Type II GRBs, $k = 0.39^{+0.02}_{-0.02}$, $b = -17.82^{+0.93}_{-0.98}$. The results for both Type I and Type II GRBs, including the fitted Amati relations and their associated 3$\sigma$ intrinsic scatter regions, are shown in Fig.~\ref{fig:correlations}(a). These findings provide a comprehensive characterization of the energy correlations in GRBs, which serves as a reference for the physical origin of GRB types.

\noindent\textbf{MVT-duration ($T_{90}$)}: The minimum variability timescale (MVT) is defined as the shortest timescale of significant variation that exceeds statistical noise in the GRB temporal profile. It serves as a probe of both the central engine's activity and the geometric size of the emitting region \cite{2015ApJ...811...93G, 2023A&A...671A.112C}. The median rest-frame MVT (i.e., MVT/(1+z)) for type I and type II GRBs is 10 ms and 45 ms, respectively. To determine the MVT for GRB~250704B, we applied the Bayesian block algorithm to the full light curve in the GRM (15–1000 keV) and WXT (0.5–4 keV) bands, identifying the shortest block that encompasses the rising phase of a pulse. The MVT of GRB~250704B is approximately 5 ms in the gamma-ray band and about 37.5 ms in the X-ray band (Table~\ref{tab:obs_properties}). These values align more closely with the MVT distribution of type I GRBs \cite{2015ApJ...811...93G} than with that of type II GRBs, as shown in Fig.\ref{fig:EP250704a}d.

\noindent\textbf{Spectral lag–Peak luminosity}: Analysis shows that GRBs with shorter spectral lags generally exhibit higher peak luminosities \cite{2000ApJ...534..248N, 2012MNRAS.419..614U}. Type I GRBs typically deviate from the main trend defined by Type II GRBs, showing systematically smaller spectral lags at comparable luminosities. This distinction makes the spectral lag–peak luminosity relation a useful tool for phenomenological classification. Using previously established samples of Type I and Type II GRBs with known redshifts \cite{2010ApJ...711.1073U, 2022ApJ...924L..29X}, we construct the $L_{\gamma,\text{iso}}$–$\tau_z$ diagram, where $L_{\gamma,\text{iso}}$ represents the isotropic peak luminosity and $\tau_z = \tau/(1+z)$ is the rest-frame spectral lag. The relation for Type II GRBs is modeled as $\log L_{\gamma,\text{iso}} = b + k \log \tau_z$. The best-fit parameters with $1\sigma$ uncertainties are $k = -0.94^{+0.07}_{-0.26}$, $b = 54.24^{+0.49}_{-0.14}$, and $\log\sigma_{\mathrm{int}} = -1.61^{+0.29}_{-0.17}$. The best-fit relation and its $3\sigma$ intrinsic scatter region are shown in Fig.~\ref{fig:correlations}(b). For GRB~250704B, the rest-frame spectral lag between the 100–150 keV and 200–250 keV energy bands is measured as $\tau_z = 2.5_{-2.5}^{+3.5}~\text{ms}$. The corresponding isotropic peak luminosity, derived from SVOM-GRM spectral fitting and corrected for redshift, is $L_{\gamma,\text{iso}} = (1.8_{-0.07}^{+0.06}) \times 10^{52}~\text{erg~s}^{-1}$. As illustrated in Fig.~\ref{fig:correlations}(b), the position of GRB~250704B in this parameter space is clearly separated from the region occupied by Type II GRBs.

\noindent\textbf{Amplitude parameter ($f_{\rm eff}$)-duration ($T_{90}$)}: The amplitude parameter is defined as $f=F_{\rm p}/F_{\rm b}$, in which $F_{\rm p}$ and $F_{\rm b}$ are the peak flux and background flux at the same epoch, respectively. To test whether a short-duration GRB is an intrinsically short type I GRB or is from a disguised short-duration GRB caused by the tip-of-iceberg effect \cite{2014MNRAS.442.1922L}, an effective amplitude parameter of long-duration GRB is defined as $f_{\rm eff}=F^{'}_{\rm p}/F_{\rm b}$. Here, $F^{'}_{\rm p}$ is the simulated peak flux of prompt emission, which is rescaled down from the original long-duration GRB light curve until its signal above the background has a duration just shorter than 2 s. Following the calculation procedure presented in ref \cite{2014MNRAS.442.1922L}, we obtain the effective amplitude as $f_{\rm eff}=1.48\pm0.03$ for GRB~250704B in the gamma-ray band. It is consistent with the short-duration GRBs region in the $f_{\rm eff}$-$T_{90}$ diagram (Fig.\ref{fig:correlations}(c), confirming the genuinely short-duration feature of the burst.

\noindent\textbf{$\epsilon$-duration ($T_{90}$)}: A new phenomenological classification method for GRBs by adopting a new parameter $\varepsilon = E_{\rm \gamma,iso,52}/E^{\rm 5/3}_{\rm p,z,2}$ is proposed to identify the progenitor of GRBs \cite{2010ApJ...725.1965L}, where $E_{\rm \gamma,iso,52}$ is the isotropic gamma-ray energy in units of $10^{52}$ erg, and $E_{\rm p,z,2}$ is the cosmic rest-frame spectral peak energy in units of 100 keV. The $\varepsilon$ has a clear bimodal distribution (high- and low-$\varepsilon$ regions) with a division line at $\varepsilon \sim 0.03$, which corresponds to type I and type II GRB, respectively \cite{2010ApJ...725.1965L}. We calculate the $\varepsilon$ for GRB~250704B, and then plot it in $\epsilon-T_{90}$ diagram by comparing with other typical Type I and Type II GRBs which are taken from ref \cite{2010ApJ...725.1965L} (Fig.\ref{fig:correlations}(d)). We obtain $\varepsilon \sim 0.008$ for GRB~250704B which falls into the low-$\varepsilon$ region, suggesting that it is of a Type I origin.

\subsection{Afterglow Analysis.\\}
\noindent\textbf{X-ray and optical afterglows.} 
To investigate the physical processes in the afterglow, we corrected the afterglow flux using the dustmaps package \cite{2018JOSS....3..695G}. We adopted the Corrected SFD Reddening Map (CSFD) \cite{2023ApJ...958..118C} to obtain the Galactic color excess $E(B-V)$ along the line of sight. The optical photometry was then dereddened with the G23 extinction relation \cite{2023ApJ...950...86G} implemented in the dust$\_$extinction \cite{Gordon2024} before being fitted to the afterglow model.

Empirically, we fitted the X-ray and optical afterglow light curves with multi-segment broken power-laws. For the X-ray light curve, the best model by Bayesian Information Criterion (BIC) is a broken power-law with four breaks. It exhibits (i) a steep decay ($\alpha_1 = 3.3$) from $T_0 + 2.5 \times 10^{2}$~s to $6.9 \times 10^{2}$~s, (ii) a shallow decay ($\alpha_2 = 0.3$) from $6.9 \times 10^{2}$~s to $2.1 \times 10^{3}$~s, (iii) a decay ($\alpha_3 = 1.0$) from $2.1 \times 10^{3}$~s to $1.5 \times 10^{4}$~s, (iv) another shallow decay ($\alpha_4 = 0.2$) from $1.5 \times 10^{4}$~s to $7.2 \times 10^{4}$~s, and finally (v) a normal decay ($\alpha_5 = 1.6$) after $7.2 \times 10^{4}$~s. The time-integrated spectra show no significant evolution in photon index or absorption. Time-resolved spectral analysis during the steep decay phase indicates that each segment is well fit by an absorbed power-law model (Table \ref{tab:ep_spec_evo}). Assuming a common break time across all the optical and near-infrared bands, the light curves show two phases: (i) a slow rise ($\alpha = 0.1$) during the early afterglow, coincident with the X-ray shallow decay, and (ii) after $7.5 \times 10^{4}$~s a steep decay ($\alpha = 2.7$). The discrepancy between the X-ray and optical light curves cannot be adequately explained by a standard afterglow model, suggesting the presence of energy injection. The simple decaying behavior in X-ray, after $7.5 \times 10^{4}$~s, is consistent with emission dominated by a normal forward shock.

There are two possible interpretations for the late-stage ($>10^4$~s post-burst) optical afterglow: an external origin or an internal origin. Based on the prompt-emission properties and the absence of a supernova-like component in the late-time optical afterglow, we adopted a uniform-density interstellar medium (ISM) and an on-axis viewing angle of $0^\circ$ in both scenarios.

For the external-origin scenario, we computed the forward-shock light curve with the numerical code PyFRS (\url{https://github.com/leiwh/PyFRS}) \cite{2013NewAR..57..141G, 2018pgrb.book.....Z, 2023ApJ...948...30Z}, assuming a top-hat jet. Posterior distributions were obtained via Markov-chain Monte Carlo (MCMC) sampling implemented in the Python package emcee \cite{2013PASP..125..306F}. We employed 70 walkers evolving for 30000 steps, with the first 18000 steps discarded as the burn-in phase. The resulting parameters are listed in Table \ref{tab:parameters_distributions_FS}, and the corresponding best-fit model light curves are presented in Fig. \ref{fig:external_mcmc}. The MCMC fitting yields a reasonable jet energy and a low circumburst density, but the initial Lorentz factor $\Gamma_0$ and the jet opening angle $\theta_0$ are both very small, making an external-shock origin difficult to reconcile with the data.

For the internal-origin scenario, the central engine must have remained active for a very long time, so that the multi-wavelength afterglow at late times is still dominated by radiation related to central-engine activity. This interpretation is supported by the multi-wavelength light curves: the optical decay is steeper than the X-ray decay, while the late-time X-ray emission alone follows a normal pre jet-break afterglow. In this scenario, it is difficult to strongly constrain the external-shock model parameters because the external-shock emission is largely buried. Nevertheless, we can still place some simple constraints based on the basic properties of the external-shock model. The X-ray spectral index $\beta \sim 1.3$ and temporal slope $\alpha \sim 1.6$ satisfy the standard closure relations for the pre jet-break phase ($\beta = p/2$ and $\alpha = (3\beta-1)/2 \sim 1.5$), indicating that the X-rays have not yet entered a jet break. However, both the optical and X-ray light curves appear to be contaminated by early internal activities, which prevents us from reliably identifying the afterglow onset peak and from constraining the initial Lorentz factor. In addition, since the late-time X-rays show no sign of a jet break, constraining the jet opening angle $\theta_j$ is also difficult. But we could still estimate meaningful lower limits from the available data. Because the optical afterglow onset peak was not detected, the peak time $t_{\rm p}$ is only constrained to be earlier than $2\times10^4\rm \ s$, beyond which the X-ray emission is dominated by the afterglow emission. Substituting this upper limit into the standard relation \cite{1999ApJ...520..641S}:
\begin{equation}
\Gamma_0 = 193\,(n\,\eta)^{-1/8}\left(\frac{E_{\rm \gamma,iso}/10^{52}~\rm erg}{t_{\rm p,z}/100~\rm s}\right)^{1/8},
\end{equation}
where $n$ is the ISM density which $\sim 0.1$ for Type I GRBs and $\eta \equiv E_{\rm \gamma,iso}/E_{\rm K,iso} \sim 0.2$ is the prompt-emission efficiency, yields a lower limit on $\Gamma_0$. Similarly, the lower limit of jet break time should not be smaller than 1.2 day, which leads to a lower limit on the jet opening angle via \cite{1999ApJ...525..737R,1999ApJ...519L..17S,2001ApJ...562L..55F}
\begin{equation}
\theta_{\rm j}=0.07~{\rm rad} \left(\frac{t_{\rm break}}{1\ \rm day}\right)^{3/8} \left(\frac{1+z}{2}\right)^{-3/8} \left(\frac{E_{\rm K,iso}}{10^{53}\ \rm erg}\right)^{-1/8}\left(\frac{n}{0.1\ \rm cm^{-3}}\right)^{1/8}.
\end{equation}
Adopting typical parameters, we can therefore place approximate lower limits of $\Gamma_0 > 80$ and $\theta_{\rm j} > 4.3^\circ$, respectively.

\noindent\textbf{Short GRBs and X-ray afterglow comparison.} 
Using Swift-detected short GRBs and their afterglow samples, we perform a comparative analysis for each component of EP250704a, including the main spike (detected predominantly in gamma rays), the extended emission (observed only in X-rays), and the X-ray afterglow. We specifically focus on short GRBs exhibiting extended emission during the prompt phase, as well as those showing internal plateaus or shallow decays in the afterglow.
\begin{itemize}
    \item \textbf{Short GRBs with extended emission.} As of 2018, Swift has detected $\sim1000$ GRBs, among which 40\% have measured redshifts. In many cases, short GRBs with extended emission have a BAT $T_{90}$ longer than 2 s. To identify such events, we construct cumulative light curves for all GRBs with known redshifts. A GRB is classified as an short GRB with extended emission if its signal duration obviously longer than its $T_{90}$ and have recognized extended emission component by both BAT catalog and at least one reference. Here we obtain 9 samples (confirmed by ref \cite{2020RAA....20..201Z}). For comparison, we calculate the luminosity (15--150 keV) of the extended emission and the photon index. The luminosity of EP250704a is derived from its WXT spectrum by extrapolating to the 15–150 keV energy range. As illustrated in Fig. \ref{fig:outlier}, EP250704a lies close to the median value of short GRBs redshift distribution and surpasses 53\% of the short GRB sample. However, when compared specifically with short GRBs that exhibit extended emission, the soft X-ray component of EP250704a appears as an outlier: it has a relatively low extended emission luminosity with soft photon index and a relatively high redshift (exceeding 70\% of short GRBs with extended emission). These properties make its extended emission component difficult to detect by instruments operating in higher energy bands, supporting the view that such soft X-ray emission may be more common but was previously missed due to instrumental limitations. We note that their prevalence within the short GRB population remains to be established with a larger sample of events.
    \item \textbf{The shallow decays.} We download afterglow data for 270 short GRBs. Among these, 244 have X-ray afterglow detections, and 169 contain sufficient data points for modeling. We fit the afterglow light curves using a multi-segment broken power-law model with the multivariate adaptive regression spline (MARS) fitting tool\footnote{\url{https://pypi.org/project/py-earth2/}}. After excluding afterglows exhibiting strong X-ray flares, we define a shallow decay segment as having a time span larger than 0.4 dex (log scale) and a temporal slope shallower than -0.8 \cite{2019ApJ...883...97Z}. This selection yields 18 short GRBs, as also reported in the literature \cite{2019ApJS..245....1T}.
    \item \textbf{The internal plateaus.} Using the same X-ray afterglow sample, we search for internal plateaus by adopting a modified criterion: the plateau must have a time span between 0.1 dex and 0.4 dex (log scale), a slope shallower than -0.8, and followed by a steep decay with slope steeper than $\Gamma_X+1$. This selection identifies 8 short GRBs, consistent with previously published samples \cite{2019ApJS..245....1T}.
\end{itemize}

\noindent\textbf{Non-detection of supernova.} 
Deep multi-band imaging was carried out with the GTC 3 to 10 days after $T_{0}$ and with VLT/FORS2 12 and 20 days after $T_{0}$ in $i$-band. Here we compared the photometric upper limits obtained from these late time observations with the light curves of well studied and observed supernovae. We used the data of: SN2025kg \cite{Li_2025}, SN1998bw \cite{Clocchiatti_2011} and SN2006aj \cite{Sollerman_2006}. We applied the cosmological correction, due to the expansion of the Universe, both to the time of observation, to go from rest to observer frame, and to the magnitudes, to account for the different distances of the afterglow and the supernovae emissions. The comparison reveals no evidence of an associated supernova (Fig. \ref{fig:afterglow}). Furthermore, empirical fitting of the late-time multi-band afterglow shows a simple power-law decay with an index of $\sim 2$ and no signature of rebrightening, further supporting the lack of a supernova component.

\noindent\textbf{Host galaxy.}
From the optical/nIR afterglow, we infer only a small amount of reddening at the GRB site, i.e., $E(B-V)=0.12\pm0.04$ mag, corresponding to $A_V\approx0.5$ mag for a Milky Way-like extinction law, thus disfavoring a heavily dust-obscured environment. A more stringent constraint comes from the optical limits on the host-galaxy brightness. We derive an $r$-band limit of $r>25.37$ AB, corresponding to $M_u \gtrsim -18$ mag at $z=0.66$, since the observed $r$ band probes the rest-frame near-UV/$u$ band. Using the relation of \cite{Hopkins2003}, converted to a Chabrier IMF \cite{Chabrier2003}, and adopting the inferred intrinsic extinction of $A_V\approx0.5$ mag, we obtain ${\rm SFR}<0.16~M_\odot~{\rm yr}^{-1}$. Therefore, either the GRB was hosted by a galaxy with a low star-formation rate, if $A_{V,\rm gal}\approx A_{V,\rm GRB}$, or, if most of the star formation is dust-obscured ($A_{V,\rm gal}\gg A_{V,\rm GRB}$), the burst occurred far from the main star-forming regions.

\noindent\textbf{Detection rate estimation from EP.}
A simple estimate from EP detections suggests that short GRBs with long-lasting soft X-ray prompt emission occur at a rate of several tens per year over the full sky. Since the launch of EP in January 2024, two such events, EP250704a and EP260527a \cite{EP260527aGCN}, have been discovered, corresponding to an observed detection rate of $\sim$ 1 yr$^{-1}$. Taking into account the effective EP observing duty cycle of about 50\% and the WXT instantaneous sky coverage of $\sim$ 1/11, this implies an all-sky event rate of roughly 22 yr$^{-1}$ for short GRBs with long-lasting soft X-ray prompt emission detectable by EP-WXT.

\clearpage

\clearpage

\setcounter{figure}{0}
\renewcommand{\thefigure}{S\arabic{figure}}
\captionsetup[figure]{name={\bf Figure}, labelsep=period}

\setcounter{table}{0}
\renewcommand{\thetable}{S\arabic{table}}
\captionsetup[table]{name={\bf Table}, labelsep=period}

\clearpage
\begin{sidewaystable*}
\centering
\tiny
\caption{Spectral fitting results and corresponding fitting statistics. All errors represent the 1$\sigma$ uncertainties.}
\label{tab:spike_spec_fit}
\begin{tabular}{ccccccccccccccc}
\hline
\hline
\multirow{3}{*}{t1 (s)} & \multirow{3}{*}{t2 (s)} & \multicolumn{5}{c}{CPL model} & & \multicolumn{7}{c}{CSBPL model}\\
\cline{3-7}
\cline{9-15}
& & $\alpha$ & $E_{\rm p}$ & log$A$ & stat/dof & BIC & & $E_{\rm b}$ & $E_{\rm p}$ & $\alpha_1$ & $\alpha_2$ & log$A$ & stat/dof & BIC\\
& & & \multicolumn{3}{c}{($\rm{photons~cm^{-2}~s^{-1}~keV^{-1}}$)} & & & (keV) & (keV) & & \multicolumn{3}{c}{($\rm{photons~cm^{-2}~s^{-1}~keV^{-1}}$)} & \\
\hline
-0.10 & 0.10 & ${-1.27}_{-0.09}^{+0.10}$ & ${334.97}_{-35.05}^{+43.48}$ & ${-0.96}_{-0.04}^{+0.04}$ & 160.08/87 & 173.58 & 
& ${77.62}_{-61.41}^{+28.54}$ & ${272.27}_{-34.04}^{+121.28}$ & ${-1.08}_{-0.12}^{+0.67}$ & ${-1.79}_{-0.07}^{+0.47}$ & ${1.28}_{-0.77}^{+0.21}$ & 156.13/85 & 178.63\\
0.10 & 0.40 & ${-1.08}_{-0.08}^{+0.07}$ & ${626.61}_{-59.07}^{+87.88}$ & ${-1.04}_{-0.03}^{+0.03}$ & 190.26/120 & 204.69 & 
& ${58.75}_{-24.79}^{+28.75}$ & ${851.14}_{-194.99}^{+203.25}$ & ${-0.60}_{-0.25}^{+0.54}$ & ${-1.44}_{-0.15}^{+0.20}$ & ${0.38}_{-0.73}^{+0.39}$ & 182.25/118 & 206.31\\
-0.10 & 0.40 & ${-1.13}_{-0.06}^{+0.06}$ & ${489.78}_{-35.84}^{+57.24}$ & ${-0.99}_{-0.03}^{+0.02}$ & 163.85/128 & 178.48 & 
& ${55.85}_{-26.47}^{+57.13}$ & ${588.84}_{-122.18}^{+91.93}$ & ${-0.78}_{-0.27}^{+0.31}$ & ${-1.46}_{-0.21}^{+0.22}$ & ${0.75}_{-0.41}^{+0.40}$ & 156.73/126 & 181.11\\
\hline
0.10 & 0.70 & ${-1.05}_{-0.05}^{+0.06}$ & ${613.76}_{-59.14}^{+59.21}$ & ${-1.04}_{-0.03}^{+0.03}$ & 194.83/123 & 209.34 & 
& ${100.93}_{-46.35}^{+41.96}$ & ${748.17}_{-135.82}^{+328.30}$ & ${-0.93}_{-0.09}^{+0.12}$ & ${-1.52}_{-0.18}^{+0.25}$ & ${0.88}_{-0.17}^{+0.15}$ & 188.88/121 & 213.06\\
\hline
\hline
\end{tabular}
\end{sidewaystable*}

\begin{figure}
\centering
\includegraphics[width=0.7\linewidth]{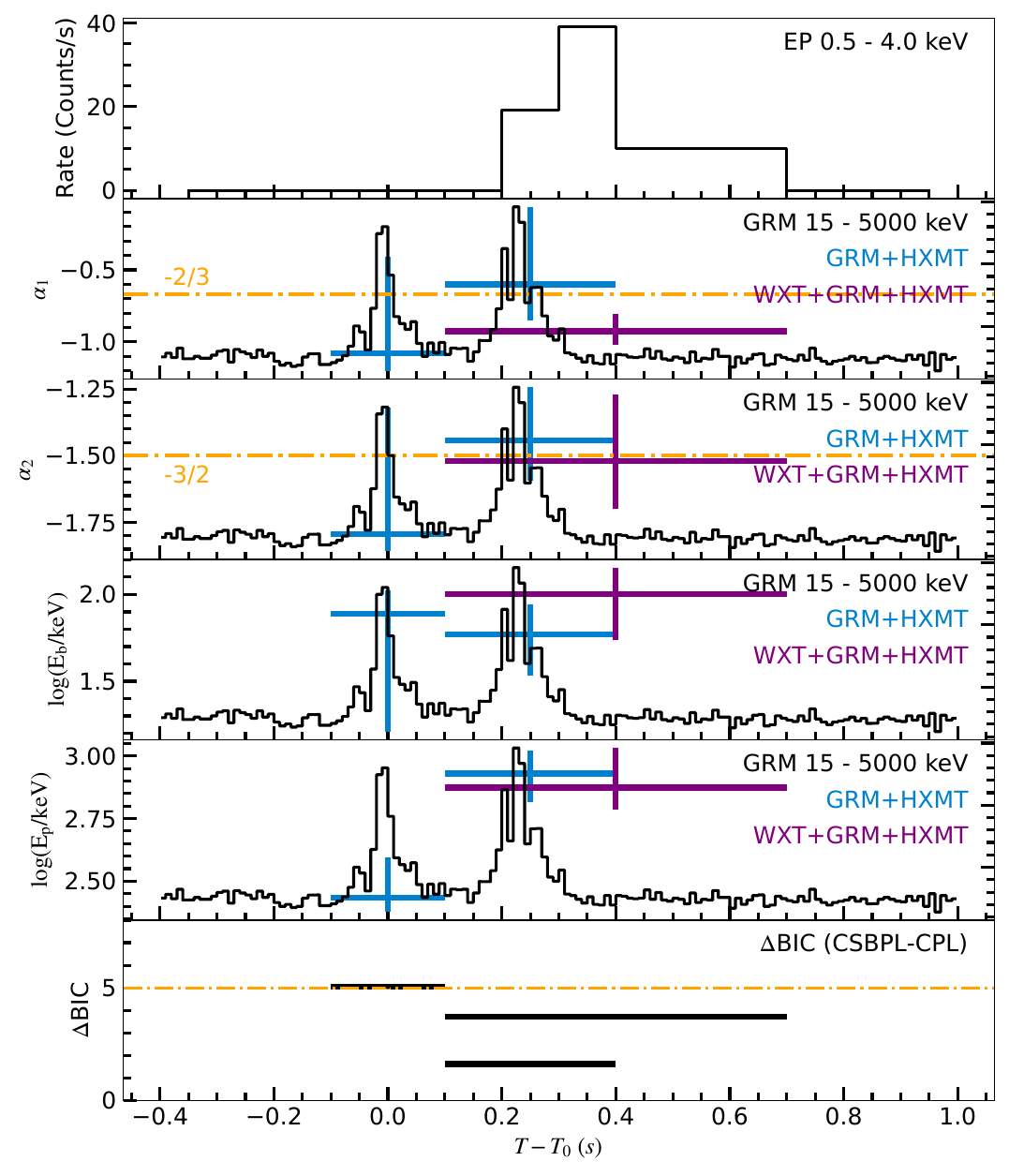}
\caption{\noindent\textbf{Spectral evolution based on the best-fit parameters of CSBPL model and corresponding statistics.} The blue points represent the best-fit parameters derived from the joint GRM and HXMT fits, while the purple points are obtained from the joint GRM, HXMT and WXT fits. The bottom panel shows the BIC differences between the best-fit CSBPL and PL models. All error bars mark the 1$\sigma$ confidence level. }
\label{fig:HE_spec_evo}
\end{figure}

\begin{figure}
\centering
\includegraphics[width=0.5\linewidth]{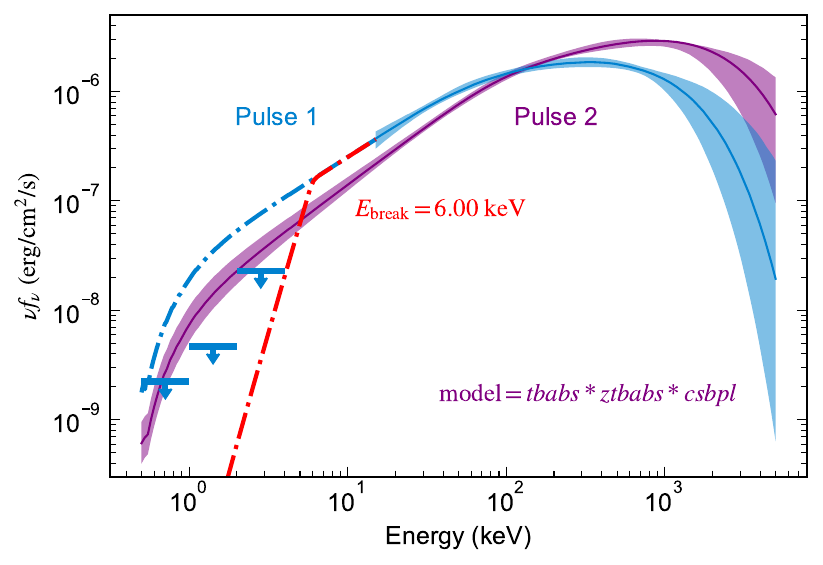}
\caption{\noindent\textbf{SEDs of the first and second pulses in the initial spike.} The SEDs are obtained from the spectral fittings with the absorbed CSBPL model. The upper limits are constrained by the WXT non-detection of the first pulse. The synchrotron self-absorption break is fixed at 6.00 keV, and the low-energy spectral slope is fixed at 3.}
\label{fig:spike_sed}
\end{figure}

\begin{figure}
    \centering
    \includegraphics[width=0.5\linewidth]{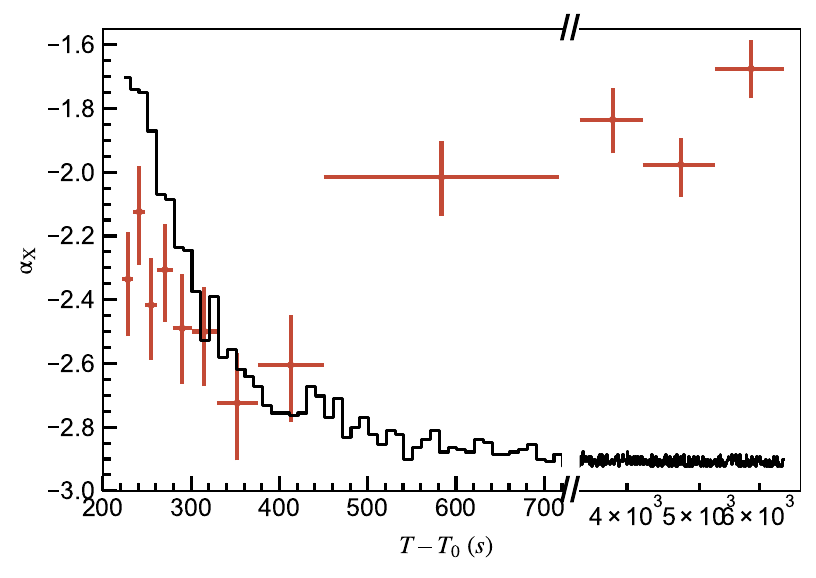}
    \caption{\noindent\textbf{The evolution of spectral index $\alpha_{\rm X}$.} The black curve represents the FXT light curve of EP250704a. The red points represent the best-fit spectral indices in the time-resolved spectral fittings with fixed $N_{\rm H}$. All error bars on data points represent their 1$\sigma$ confidence level.}
    \label{fig:fxt_spec_evo}
\end{figure}

\begin{figure}
    \centering
    \includegraphics[width=0.5\linewidth]{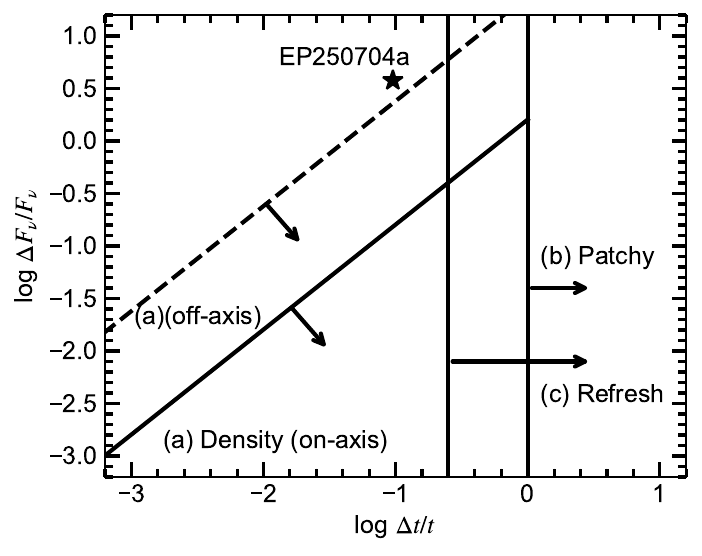}
    \caption{\noindent\textbf{The variability plane of relative timescale and flux.} Three different theoretical limits are presented: (a) density fluctuations (for both on- and off-axis case), (b) patchy shells, and (c) refreshed shocks. The variability of Episode III of EP250704a is shown by a solid star that violates all of these limits.}
    \label{fig:variability_plane}
\end{figure}

\begin{figure}
    \centering
    \includegraphics[width=0.5\linewidth]{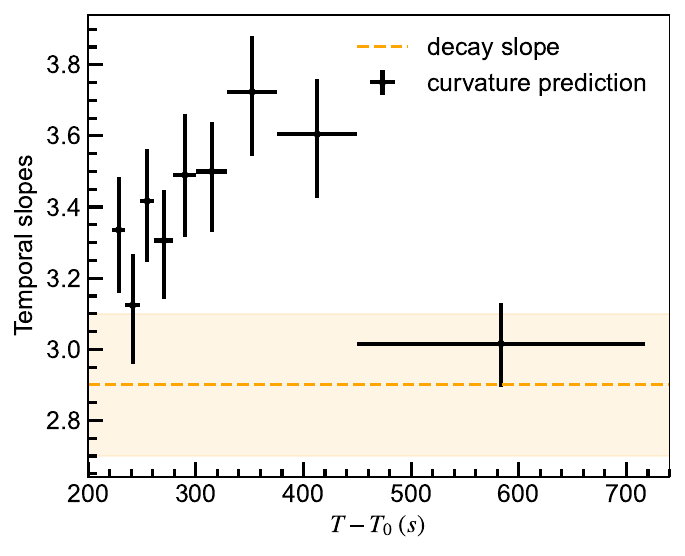}
    \caption{\noindent\textbf{The measured and curvature effect predicted temporal slopes of the steep decay phase of Episode III.} The dashed line represents the temporal slope of the steep decay phase of Episode III, while the data points depict the temporal slopes predicted by the curvature effect. The error bars on the data points represent the 1$\sigma$ confidence level.}
    \label{fig:curvature}
\end{figure}

\begin{figure}
    \centering
    \includegraphics[width=0.4\textwidth]{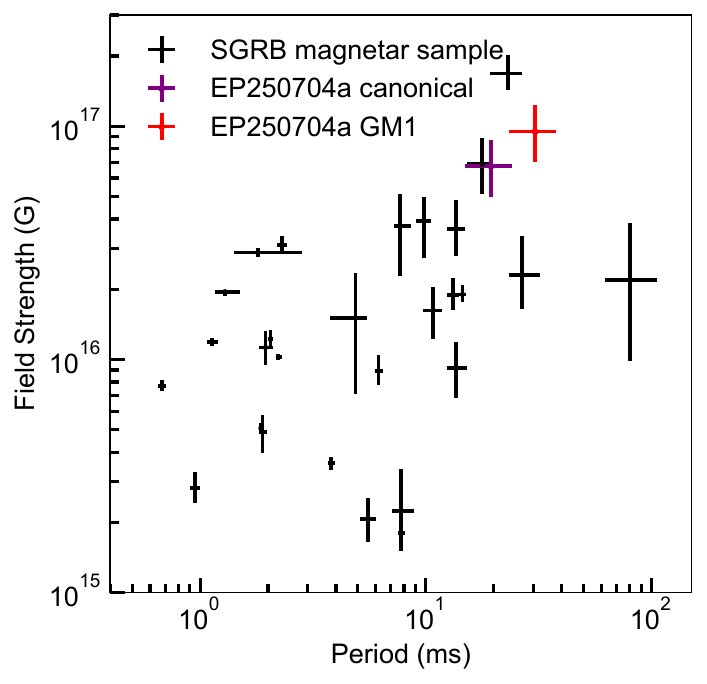}
    \includegraphics[width=0.4\textwidth]{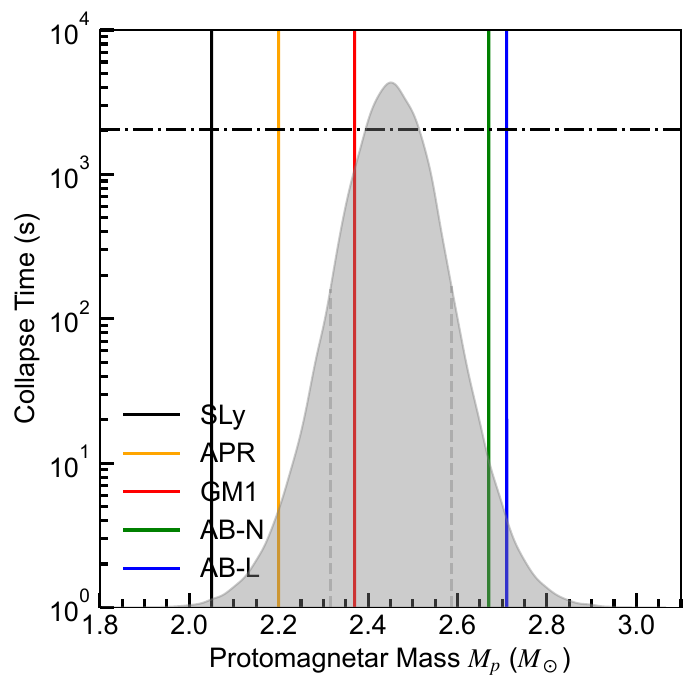}
    \caption{\noindent\textbf{Magnetar parameters derived from short GRBs and collapse time as a function of the protomagnetar mass for EP250704a.} \textit{Left:} The black data points are adopted from the surface magnetic field strength and initial spin period inferred from short GRB magnetars \cite{2013MNRAS.430.1061R}. The purple and red data points are derived from canonical and GM1 EOS magnetar mass and radius, respectively. \textit{Right:} The colored curves represent the collapse time and protomagnetar mass relation of different EOS models. The dashed and dotted horizontal line denotes the assumed collapse time of EP250704a. The shaded region in the background shows the protomagnetar mass distribution derived from the Galactic NS-NS binary systems with the 68\% mass intervals illustrated with the vertical dashed lines. The error bars on the colored data points represent the 1$\sigma$ confidence level.}
    \label{fig:magnetar_params}
\end{figure}

\clearpage
\begin{figure}[htbp]
    \begin{minipage}[b]{0.49\textwidth}
        \begin{overpic}[width=\linewidth]{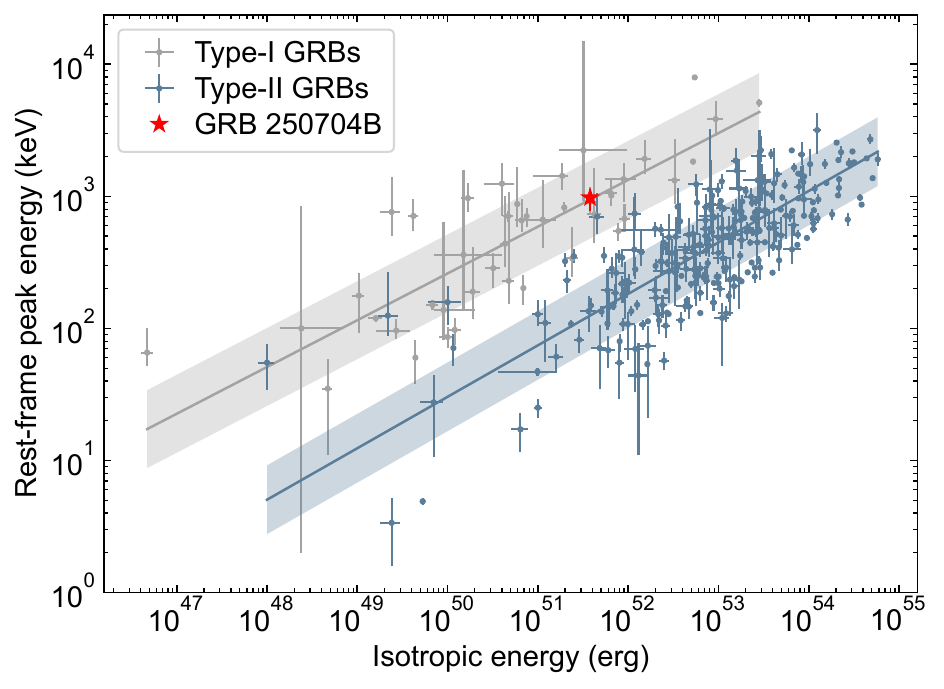}
            \put(-1,70){\textbf{\textsf{(a)}}}
        \end{overpic}
    \end{minipage}
    \hfill
    \begin{minipage}[b]{0.49\textwidth}
        \begin{overpic}[width=\linewidth]{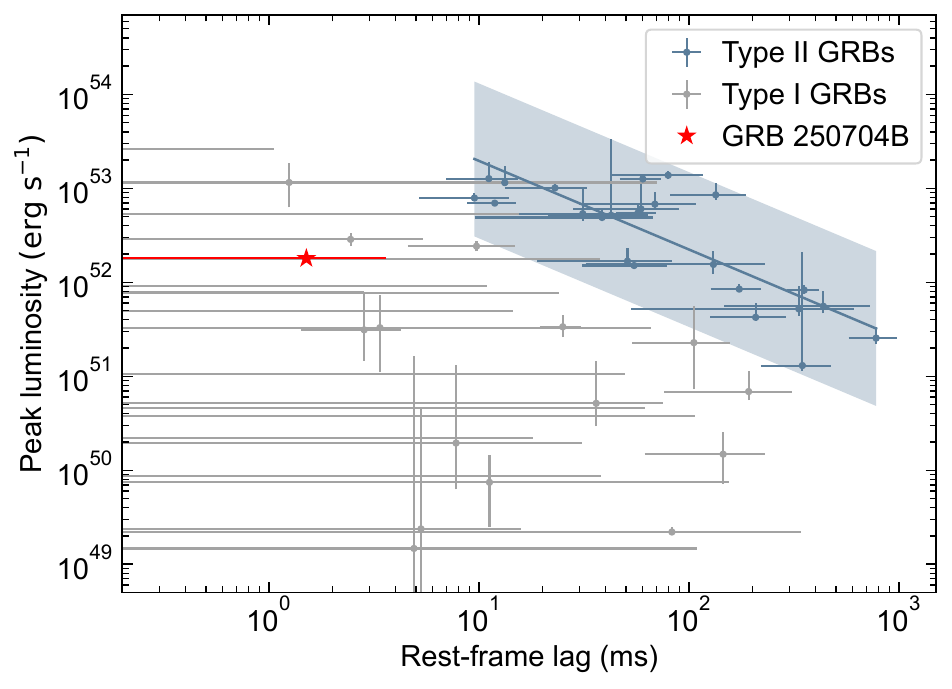}
            \put(-1,70){\textbf{\textsf{(b)}}}
        \end{overpic}
    \end{minipage}
    \vspace{0.1cm}
    \begin{minipage}[b]{0.49\textwidth}
        \begin{overpic}[width=\linewidth]{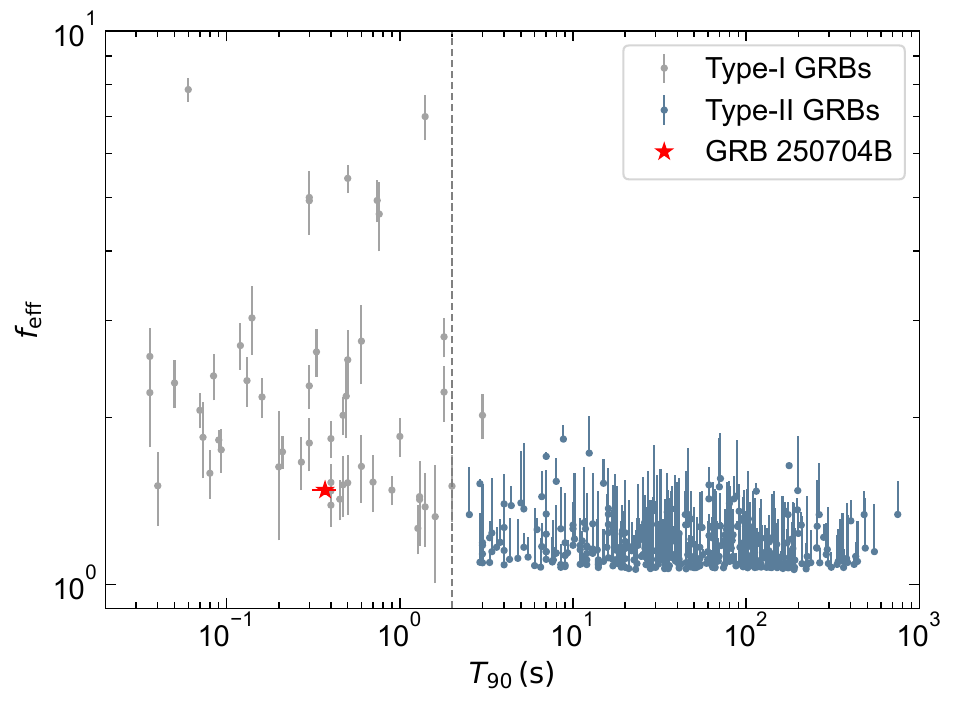}
            \put(-1,70){\textbf{\textsf{(c)}}}
        \end{overpic}
    \end{minipage}
    \hfill
    \begin{minipage}[b]{0.49\textwidth}
        \begin{overpic}[width=\linewidth]{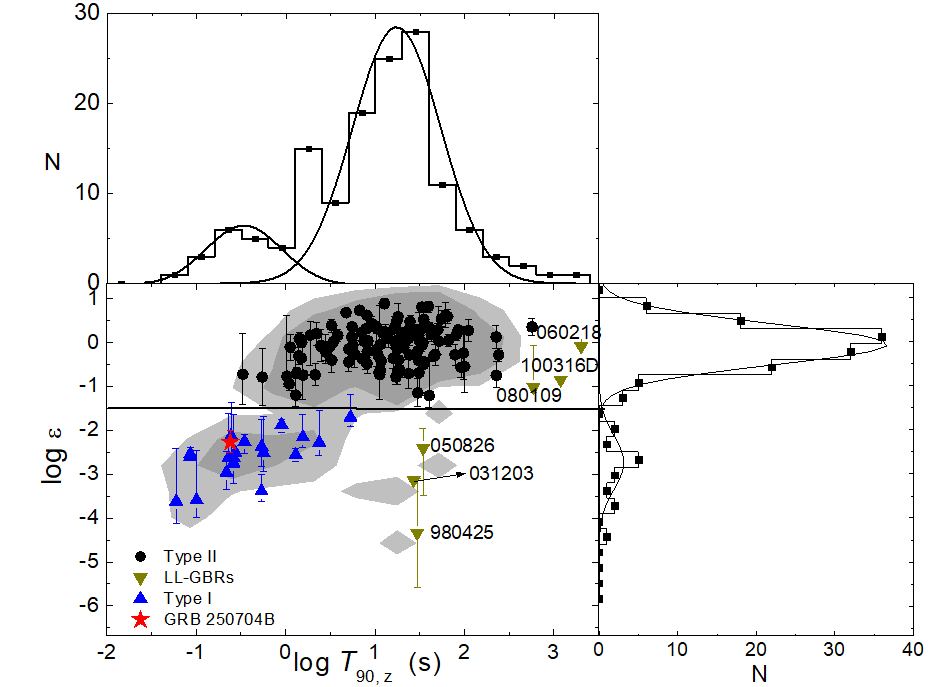}
            \put(-1,70){\textbf{\textsf{(d)}}}
        \end{overpic}
    \end{minipage}

    \caption{\textbf{The relationship of the observation properties from prompt emission of EP250704a/GRB~250704B.}
    (a) The Rest-frame peak energy versus T90 diagram (Amati relation). Type I and type II GRBs are represented by grey and cadet blue solid circles, respectively. EP250704a/GRB 250704B is highlighted by red star. 
    (b) The peak luminosity versus spectral lag diagram. Type I and type II GRBs are represented by grey and cadet blue solid circles, respectively. EP250704a/GRB 250704B is highlighted by red star. 
    (c) The $f_{eff}$ versus $T_{90}$ diagram. Type I and type II GRBs are represented by grey and cadet blue solid circles, respectively. EP250704a/GRB 250704B is highlighted by red star. 
    (d) The $\epsilon$ versus $T_{90}$ diagram. Type I, type II and low luminosity GRBs are represented by blue triangles, black circles and green triangles, respectively. EP250704a/GRB 250704B is highlighted by red star.}
    \label{fig:correlations}
\end{figure}

\begin{figure}
    \centering
    \includegraphics[width=0.9\linewidth]{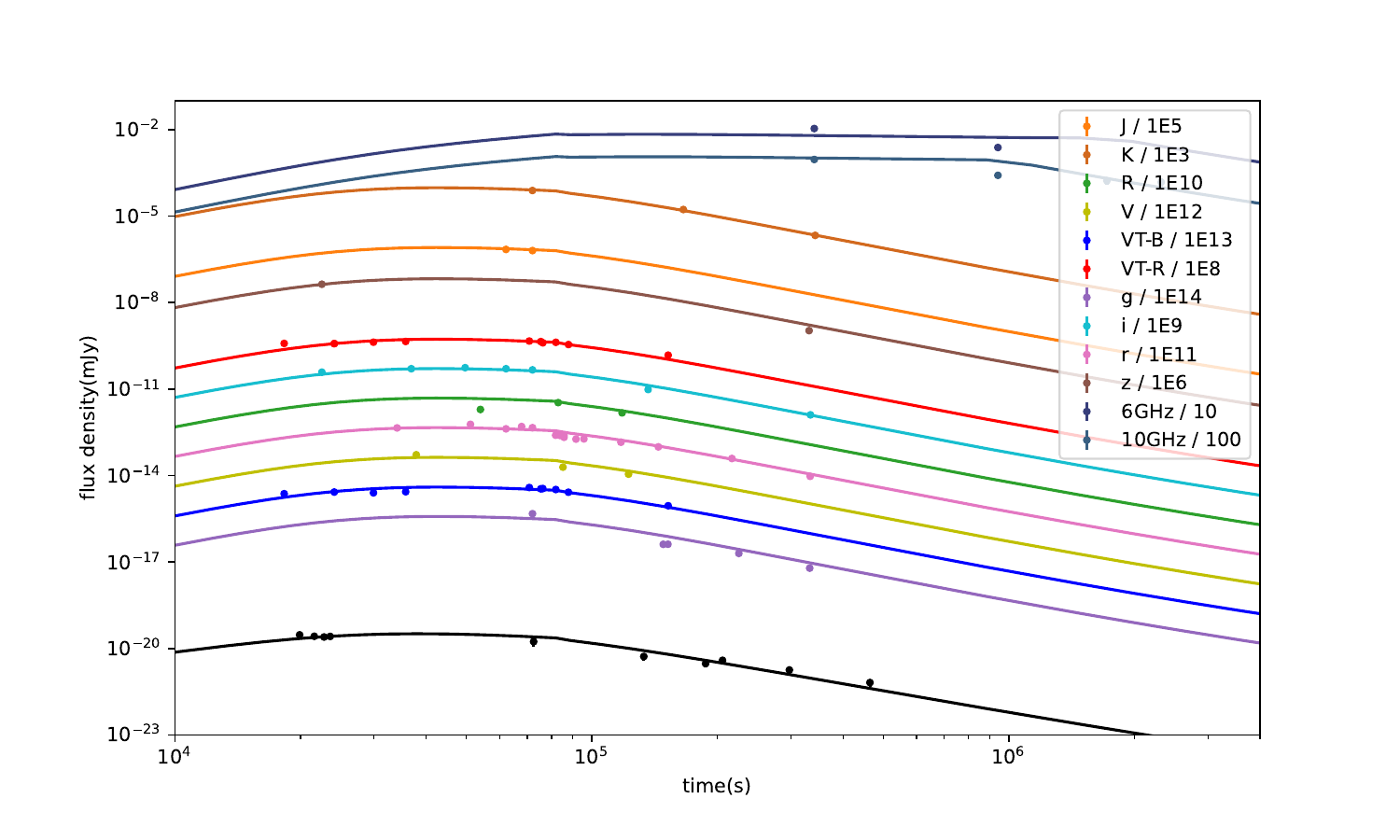}
    \caption{\textbf{MCMC modeling of the late-time multiwavelength afterglow.} Assuming that the afterglow emission after $10^4$ s originates only from the external forward shock, the solid curves show the best-fit light curves derived from the posterior parameter values listed in Table \ref{tab:parameters_distributions_FS}.}
    \label{fig:external_mcmc}
\end{figure}

\begin{figure}
    \centering
    \includegraphics[width=0.9\linewidth]{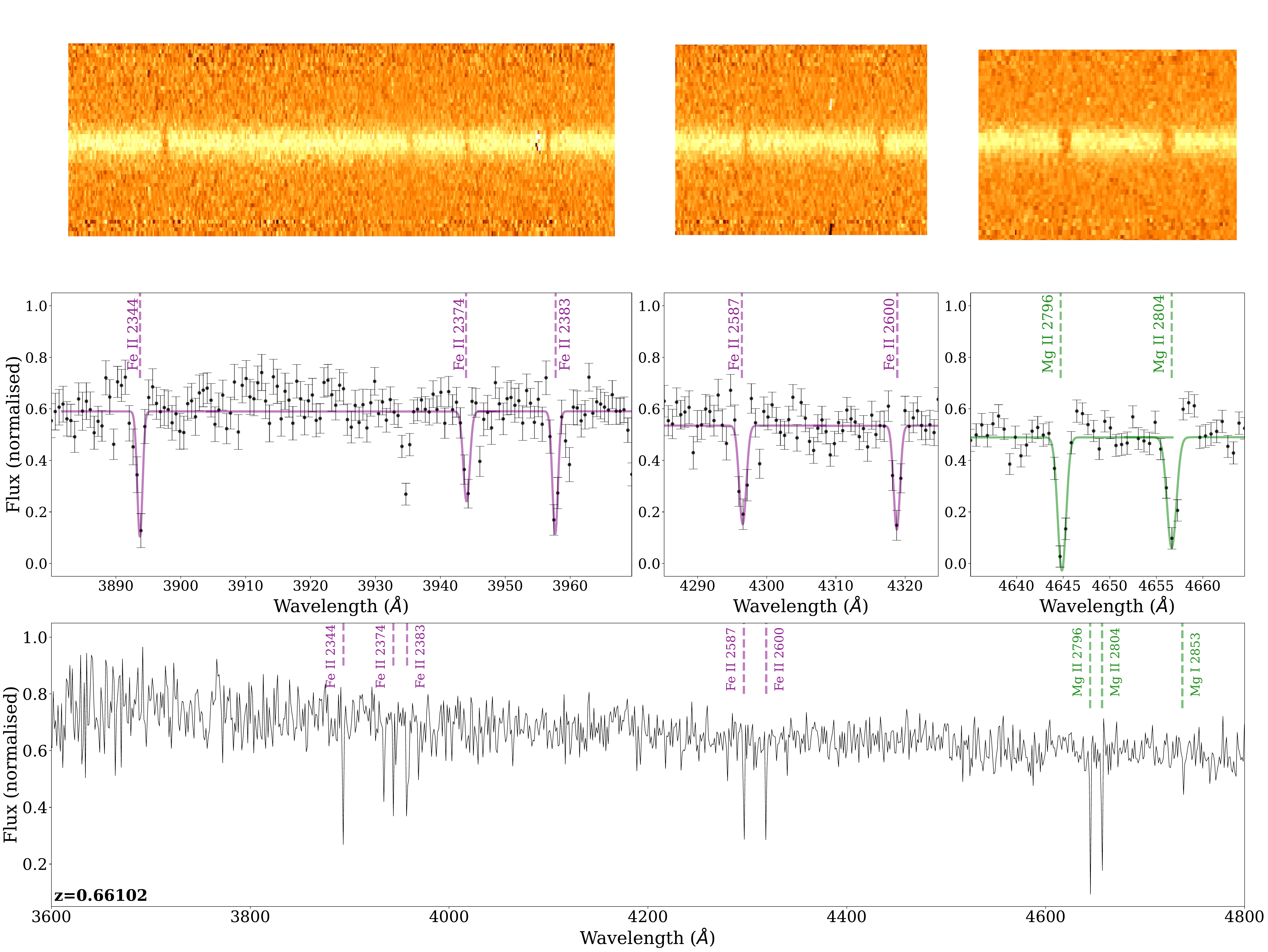}
    \caption{\noindent\textbf{Afterglow spectrum of EP250704a.} Multiple absorption lines at a common redshift of $z$=0.66102 are imprinted on the afterglow light.
    \label{fig:spec_xshooter}}
\end{figure}

\begin{figure}
    \centering
    \includegraphics[width=0.45\linewidth]{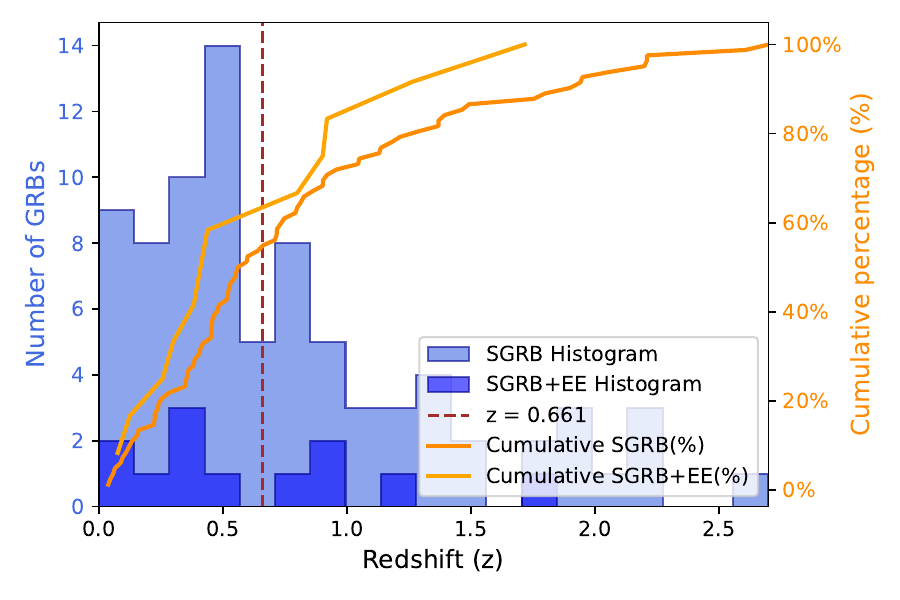}
    \includegraphics[width=0.45\linewidth]{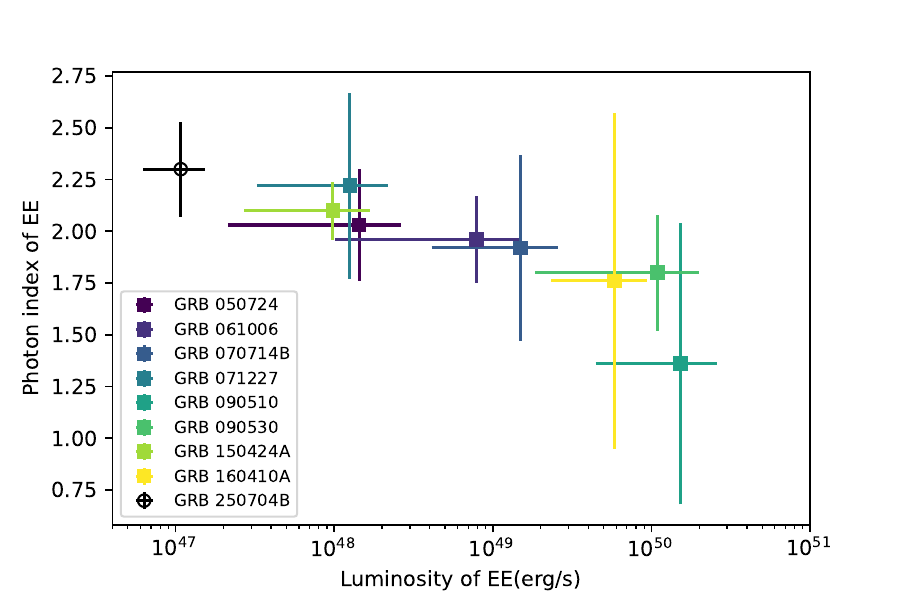}
    \caption{\textbf{Comparison of EP250704a with cosmological short GRBs}. \textit{Left panel}: The redshift distribution and accumulative curve of short GRBs and short GRBs with extended emission are presented. The red vertical dashed line is the redshift of EP250704a. \textit{Right panel}: The data points denote the luminosity and photon index of the short GRB extended emissions. The luminosity is derived in the energy range of 15--150 keV.}
    \label{fig:outlier}
\end{figure}

\begin{table}
\centering
\scriptsize
\caption{Spectral fitting results and corresponding fitting statistics for EP/WXT and EP/FXT. All errors represent the 1$\sigma$ uncertainties.}
\label{tab:ep_spec_evo}
\begin{tabular}{cccccc}
\hline
\hline
\multirow{2}{*}{$t1$} & \multirow{2}{*}{$t2$} & \multicolumn{4}{c}{PL Model}\\
\cmidrule(lr){3-6}
& & $\alpha_{X}$ & log$A$ & pgstat/dof & BIC\\
(s) & (s) & &($\rm{photons~cm^{-2}~s^{-1}~keV^{-1}}$) & & \\
\hline
1.50 & 3.50 & ${-1.05}_{-0.97}^{+0.88}$ & ${-1.80}_{-1.85}^{+1.61}$ & 0.19/1 &  2.39 \\
22.76 & 71.77 & ${-2.64}_{-0.66}^{+0.62}$ & ${-5.83}_{-1.36}^{+1.22}$ & 4.28/4 &  7.87 \\
71.77 & 115.29 & ${-1.99}_{-0.63}^{+0.49}$ & ${-4.45}_{-1.26}^{+0.92}$ & 5.87/3 &  9.08 \\
115.29 & 160.80 & ${-2.07}_{-0.40}^{+0.32}$ & ${-4.64}_{-0.78}^{+0.62}$ & 12.28/10 &  17.25 \\
160.80 & 764.50 & ${-2.33}_{-0.39}^{+0.29}$ & ${-6.08}_{-0.78}^{+0.56}$ & 11.11/16 &  16.89 \\
22.76 & 764.50 & ${-2.30}_{-0.23}^{+0.23}$ & ${-5.75}_{-0.46}^{+0.46}$ & 174.26/39 &  181.68 \\
\hline
222.00 & 234.50 & ${-2.34}_{-0.18}^{+0.15}$ & ${-5.65}_{-0.34}^{+0.28}$ & 58.83/57 & 66.98\\
234.50 & 248.00 & ${-2.12}_{-0.17}^{+0.14}$ & ${-5.29}_{-0.31}^{+0.27}$ & 49.25/58 & 57.43\\
248.00 & 261.50 & ${-2.42}_{-0.17}^{+0.15}$ & ${-5.86}_{-0.32}^{+0.28}$ & 51.46/56 & 59.58\\
261.50 & 279.00 & ${-2.31}_{-0.16}^{+0.14}$ & ${-5.75}_{-0.30}^{+0.26}$ & 40.87/55 & 48.96\\
279.00 & 300.50 & ${-2.49}_{-0.17}^{+0.17}$ & ${-6.21}_{-0.34}^{+0.33}$ & 45.98/56 & 54.1\\
300.50 & 329.00 & ${-2.50}_{-0.17}^{+0.14}$ & ${-6.34}_{-0.33}^{+0.25}$ & 62.13/56 & 70.25\\
329.00 & 375.50 & ${-2.72}_{-0.18}^{+0.16}$ & ${-6.97}_{-0.35}^{+0.30}$ & 49.12/56 & 57.25\\
375.50 & 450.00 & ${-2.60}_{-0.18}^{+0.16}$ & ${-7.12}_{-0.36}^{+0.30}$ & 68.94/55 & 77.03\\
450.00 & 717.00 & ${-2.02}_{-0.12}^{+0.11}$ & ${-6.36}_{-0.24}^{+0.21}$ & 65.42/78 & 74.18\\
3460.00 & 4196.50 & ${-1.83}_{-0.11}^{+0.10}$ & ${-6.67}_{-0.20}^{+0.18}$ & 122.06/107 & 131.44\\
4196.50 & 5234.50 & ${-1.98}_{-0.10}^{+0.08}$ & ${-7.00}_{-0.20}^{+0.16}$ & 129.85/126 & 139.55\\
5234.50 & 6464.00 & ${-1.68}_{-0.09}^{+0.09}$ & ${-6.45}_{-0.18}^{+0.17}$ & 151.30/145 & 161.28\\
\hline
\hline
\end{tabular}
\end{table}

\begin{ThreePartTable}
\begin{TableNotes}
\footnotesize
\item[*] All magnitude except clear band in this table were corrected by galactic absorption described in Method and reported in the AB system.
\end{TableNotes}
\begin{longtable}{cccccc}
\caption{Optical and Radio Afterglow Observations of GRB~250704B/EP250704a} \label{tab:optical_obs} \\
\toprule
\midrule   
\multicolumn{6}{c}{\textbf{Optical}}\\
\midrule
\midrule
\textbf{Time (days)} & \textbf{Observatory} & \textbf{Band} & \textbf{Magnitude\tnote{*}} & \textbf{Error} & \textbf{Reference} \\
\midrule
\endfirsthead
\midrule
\textbf{Time (days)} & \textbf{Observatory} & \textbf{Band} & \textbf{Magnitude\tnote{*}} & \textbf{Error} & \textbf{Reference} \\
\midrule
\endhead
\bottomrule
\multicolumn{6}{r}{{next}} \\
\endfoot
\bottomrule
\insertTableNotes
\endlastfoot
0.011 & BOOTES-5/JGT & clear\tnote{*} & 20.10 & 0.20 & This work \\
0.014 & COLIBRI & i & 20.24 & 0.20 & GCN 40942 \\
0.035 & VLT/FORS2 & z & 19.95 & 0.02 & GCN 40945 \\
0.260 & Pan-STARRS & i & 19.98 & 0.03 & GCN 40958 \\
0.260 & Pan-STARRS & z & 19.84 & 0.07 & GCN 40958 \\
0.395 & GROWTH-India & r & 19.84 & 0.08 & GCN 40962 \\
0.575 & GROWTH-India & i & 19.58 & 0.09 & GCN 40962 \\
0.592 & GROWTH-India & r & 19.53 & 0.09 & GCN 40962 \\
0.625 & SAO & R & 20.68 & 0.18 & GCN 40963 \\
0.658 & NOT & i & 19.56 & 0.05 & GCN 40971 \\
0.720 & FTW & i & 19.68 & 0.10 & GCN 40974 \\
0.720 & FTW & J & 19.31 & 0.20 & GCN 40974 \\
0.720 & FTW & r & 19.92 & 0.10 & GCN 40974 \\
0.833 & GSP & g & 19.76 & 0.20 & GCN 40975 \\
0.833 & GSP & i & 19.78 & 0.20 & GCN 40975 \\
0.833 & GSP & r & 19.82 & 0.20 & GCN 40975 \\
0.439 & NOWT & V & 19.60 & 0.16 & GCN 40999 \\
1.367 & Mondy-AZT-33IK & R & 20.78 & 0.05 & GCN 41024 \\
1.648 & NOT & i & 21.34 & 0.14 & GCN 40993 \\
0.833 & VLT/HAWK-I & J & 19.36 & 0.01 & This work \\
0.833 & VLT/HAWK-I & Ks & 19.18 & 0.01 & This work \\
0.833 & VLT/FORS2 & r & 19.82 & 0.01 & This work \\
0.833 & VLT/FORS2 & z & 19.66 & 0.01 & This work \\
1.917 & VLT/HAWK-I & Ks & 20.83 & 0.03 & This work \\
1.917 & VLT/FORS2 & r & 21.98 & 0.02 & This work \\
1.917 & VLT/FORS2 & z & 20.91 & 0.04 & This work \\
3.970 & VLT/HAWK-I & Ks & 23.07 & 0.09 & This work \\
3.970 & VLT/FORS2 & r & 24.07 & 0.18 & This work \\
3.970 & VLT/FORS2 & z & 23.75 & 0.19 & This work \\
12.830 & VLT/FORS2 & i & \textgreater24.98 & - & This work \\
19.896 & VLT/FORS2 & i & \textgreater24.98 & - & This work \\
19.896 & VLT/FORS2 & z & \textgreater24.64 & - & This work \\
0.212 & SVOM/VT & VT-B & 20.48 & 0.01 & This work \\
0.212 & SVOM/VT & VT-R & 19.94 & 0.01 & This work \\
0.279 & SVOM/VT & VT-B & 20.33 & 0.01 & This work \\
0.279 & SVOM/VT & VT-R & 19.96 & 0.01 & This work \\
0.346 & SVOM/VT & VT-B & 20.40 & 0.01 & This work \\
0.346 & SVOM/VT & VT-R & 19.84 & 0.01 & This work \\
0.414 & SVOM/VT & VT-B & 20.30 & 0.01 & This work \\
0.414 & SVOM/VT & VT-R & 19.77 & 0.01 & This work \\
0.819 & SVOM/VT & VT-B & 19.94 & 0.01 & This work \\
0.819 & SVOM/VT & VT-R & 19.73 & 0.01 & This work \\
0.873 & SVOM/VT & VT-B & 20.05 & 0.03 & This work \\
0.873 & SVOM/VT & VT-R & 19.78 & 0.02 & This work \\
0.882 & SVOM/VT & VT-B & 20.03 & 0.02 & This work \\
0.882 & SVOM/VT & VT-R & 19.89 & 0.02 & This work \\
0.948 & SVOM/VT & VT-B & 20.11 & 0.01 & This work \\
0.948 & SVOM/VT & VT-R & 19.85 & 0.01 & This work \\
1.016 & SVOM/VT & VT-B & 20.35 & 0.02 & This work \\
1.016 & SVOM/VT & VT-R & 20.04 & 0.01 & This work \\
1.763 & SVOM/VT & VT-B & 21.52 & 0.02 & This work \\
1.763 & SVOM/VT & VT-R & 20.96 & 0.02 & This work \\
3.211 & SVOM/VT & VT-B & \textgreater23.10 & - & This work \\
3.211 & SVOM/VT & VT-R & \textgreater23.25 & - & This work \\
4.122 & SVOM/VT & VT-B & \textgreater23.10 & - & This work \\
4.122 & SVOM/VT & VT-R & \textgreater23.25 & - & This work \\
3.838 & GTC & z & 23.86 & 0.30 & This work \\
3.851 & GTC & g & 24.48 & 0.27 & This work \\
3.862 & GTC & r & 24.04 & 0.26 & This work \\
3.868 & GTC & i & 23.68 & 0.17 & This work \\
4.794 & GTC & Ks & \textgreater22.73 & - & This work \\
4.808 & GTC & J & \textgreater23.52 & - & This work \\
4.816 & GTC & H & \textgreater23.30 & - & This work \\
9.778 & GTC & Ks & \textgreater22.68 & - & This work \\
9.793 & GTC & J & \textgreater23.12 & - & This work \\
9.802 & GTC & H & \textgreater22.95 & - & This work \\
11.658 & GTC & z & \textgreater24.27 & - & This work \\
11.671 & GTC & g & \textgreater24.92 & - & This work \\
11.681 & GTC & r & \textgreater25.37 & - & This work \\
11.688 & GTC & i & \textgreater24.82 & - & This work \\
24.682 & GTC & i & \textgreater25.31 & - & This work \\
0.427 & ALT100C & i & 19.68 & 0.05 & This work \\
1.795 & OSN & i & \textgreater20.59 & - & This work \\
0.947 & KNC & r & 20.11 & 0.22 & This work \\
0.970 & KNC & r & 20.13 & 0.15 & This work \\
0.991 & KNC & r & 20.13 & 0.12 & This work \\
0.992 & KNC & r & 20.29 & 0.32 & This work \\
1.060 & KNC & r & 20.44 & 0.26 & This work \\
1.670 & KNC & r & 21.12 & 0.23 & This work \\
1.715 & KNC & g & 21.94 & 0.24 & This work \\
1.761 & KNC & g & 21.93 & 0.22 & This work \\
2.564 & KNC & r & \textgreater21.15 & - & This work \\
2.604 & KNC & g & 22.72 & 1.12 & This work \\
0.986 & TRT-SRO & V & 20.32 & 0.16 & This work \\
1.107 & TRT-SRO & r & 20.42 & 0.09 & This work \\
1.111 & TRT-SRO & I & 20.20 & 0.15 & This work \\
1.875 & TRT-SRO & I & \textgreater20.62 & - & This work \\
2.878 & TRT-SRO & r & \textgreater21.42 & - & This work \\
1.358 & UBAI/NT-60 & r & 20.72 & 0.52 & This work \\
1.416 & UBAI/AZT-22 & V & 20.93 & 0.05 & This work \\
2.507 & UBAI/AZT-22 & r & 22.12 & 0.20 & This work \\
1.578 & KAO & i & 21.22 & 0.24 & This work \\
3.412 & AbAO-T150 & i & \textgreater22.18 & - & This work \\
\midrule
\midrule   
\multicolumn{6}{c}{\textbf{Radio}}\\
\midrule
\midrule
\textbf{Time (days)} & \textbf{Observatory} & \textbf{Band} & \textbf{Flux [$\mu Jy$]} & \textbf{Error [$\mu Jy$]} & \textbf{Reference} \\
\midrule
3.950 & VLA & 6 GHz & 109 & 8 & This work\\
3.950 & VLA & 10 GHz & 92 & 7 & This work\\
10.890 & VLA & 6 GHz & 24 & 6 & This work\\
10.890 & VLA & 10 GHz & 26 & 4 & This work\\
19.880 & VLA & 10 GHz & 17 & 5 & This work\\
26.820 & VLA & 10 GHz & 15 & 5 & This work\\
\midrule
\end{longtable}

\end{ThreePartTable}
\begin{table}
\centering
\caption{Distributions of MCMC-Fitted Parameters with PyFRS.}
\label{tab:parameters_distributions_FS}
\begin{tabular}{ccccc}
\toprule
\textbf{Parameter} & \textbf{Unit} & \textbf{Prior Type} & \textbf{Parameter Bound} & \textbf{Posterior Value}\\
\hline
$\rm log_{10}(E_{k,iso})$ & $\rm erg$ & $\rm log-uniform$ & $[51,55]$ & $52.40^{+0.03}_{-0.03}$\\
$\rm log_{10}(\Gamma_{0})$ & $1$ & $\rm log-uniform$ & $[1,3]$ & $1.46^{+0.01}_{-0.01}$\\
$\rm log_{10}(n)$ & $\rm cm^{-3}$ & $\rm log-uniform$ & $[-4,0]$ & $-2.64^{+0.07}_{-0.05}$\\
$\theta_{j}$ & $\rm degree$ & $\rm uniform$ & $[0,15]$ & $3.63^{+0.13}_{-0.07}$\\
$p$ & $1$ & $\rm uniform$ & $[2.01,2.9]$ & $2.30^{+0.01}_{-0.01}$\\
$\rm log_{10}(\epsilon_{e})$ & $1$ & $\rm log-uniform$ & $[-3,0]$ & $-0.75^{+0.02}_{-0.02}$\\
$\rm log_{10}(\epsilon_{B})$ & $1$ & $\rm log-uniform$ & $[-5,-1]$ & $-1.02^{+0.01}_{-0.03}$\\
\hline
\end{tabular}
\end{table}

\clearpage

\end{document}